\documentclass[]{article}
\usepackage[utf8]{inputenc}
\usepackage[UKenglish]{babel}
\usepackage{hyperref}
\usepackage{amsmath, bm, graphicx, subfig, enumitem, booktabs, graphicx, mathtools}
\usepackage{amsfonts}
\usepackage{algorithm}
\usepackage[noend]{algpseudocode}

\usepackage{multicol}

\usepackage{multirow, array} 
\usepackage[table]{xcolor}
\definecolor{Lightgray}{rgb}{.7,.7,.7}
\newcolumntype{A}{>{\columncolor{white}}l}
\usepackage{transparent}
\usepackage{etoolbox} 
\makeatletter
\patchcmd{\@classz}
{\CT@row@color}
{\oldCT@column@color}
{}
{}
\patchcmd{\@classz}
{\CT@column@color}
{\CT@row@color}
{}
{}
\patchcmd{\@classz}
{\oldCT@column@color}
{\CT@column@color}
{}
{}
\makeatother

\usepackage[style=authoryear, sorting=nyt, backend=biber, url=false, 
doi = false, date=year, maxcitenames=1, uniquelist=false, natbib]{biblatex}
\usepackage{csquotes}
\usepackage{fancyhdr, authblk}
\usepackage[margin=3.5cm,headheight=23pt]{geometry}

\title{Bayesian modelling for spatially misaligned health areal data: a multiple membership approach}
\author[1]{Marco Gramatica
	\footnote{Corresponding author: \href{mailto:m.gramatica@qmul.ac.uk}{m.gramatica@qmul.ac.uk}}
}
\author[2]{Peter Congdon}
\author[1,3]{Silvia Liverani}
\affil[1]{School of Mathematical Sciences, Queen Mary University of London, London, UK}
\affil[2]{School of Geography, Queen Mary University of London, London, UK}
\affil[3]{The Alan Turing Institute, The British Library, London, UK}
\date{}

\newcommand{\I}{\mathbb{I}}

\usepackage[normalem]{ulem}
\newcommand\redsout{\bgroup\markoverwith{\textcolor{red}{\rule[0.5ex]{2pt}{0.4pt}}}\ULon}

\begin{document}
	
	\maketitle
	
	\begin{abstract}
		Diabetes prevalence is on the rise in the UK, and for public health strategy, estimation of relative disease risk and subsequent mapping is important. We consider an application to London data on diabetes prevalence and mortality. In order to improve the estimation of relative risks we analyse jointly prevalence and mortality data to ensure borrowing strength over the two outcomes. The available data involves two spatial frameworks, areas (middle level super output areas, MSOAs), and general practices (GPs) recruiting patients from several areas. This raises a spatial misalignment issue that we deal with by employing the multiple membership principle. Specifically we translate areal spatial effects to explain GP practice prevalence according to proportions of GP populations resident in different areas. A sparse implementation in RStan of both the MCAR and GMCAR with multiple membership allows the comparison of these bivariate priors as well as exploring the different implications for the mapping patterns for both outcomes. The necessary causal precedence of diabetes prevalence over mortality allows a specific conditionality assumption in the GMCAR, not always present in the context of disease mapping. Additionally, an area-locality comparison is considered to locate high vs low relative risk clusters.

	\end{abstract}
	
	\section{Introduction and motivating example}

	When dealing with socio-economic or health data it is common for individuals to be classified according to nested hierarchical structures.
	
	When it is necessary to account for level specific effects, hierarchical models are better suited than standard regression techniques \parencite{Fielding2006}. However, the most common setup of such models requires lower level units to belong to only one class in the hierarchy. 
	
	As an educational example, this structure can occur for pupils who attended more than one school during their educational career; when modelling random effects for a particular outcome, it might be useful to account for the different impacts that different schools might have had on the individual student outcome. This is where multiple membership (MM) \parencite{Fielding2006} strategies become an alternative to a classical hierarchical model: by choosing appropriate weights and averaging the random effects, e.g. over different schools attended, it is possible to account for the different memberships of a specific unit at a certain level \parencite{Browne2001}. A mixed structure is also exemplified by patients that are classified by both their area of residence and the medical general practitioner (GP) primary care practice they are registered at.
	
	In this paper, the main problem under consideration is to jointly estimate relative risk of diabetes mortality and prevalence where data for the former comes from census Middle Layer Super Output Areas (MSOA), while the latter is collected for GP disease registers. Generally, in the context of areal disease mapping it is common to model outcomes (e.g. mortality, prevalence) with a generalized linear model (GLM) \parencite{Lawson2018} using area counts as observations and conditional autoregressive priors (CAR, \cite{Besag1991}) to account for spatially structured residuals. Additionally, expected disease counts or disease prevalence would be included as offsets in the model, to account for different population structures among the areas. Known risk factors are also included as covariates.
	
	This modelling strategy is based on the two outcomes sharing the same areal framework. This is not the case in the application at hand, since residents of a specific MSOA can be registered at a GP practice outside it, making it impossible to attribute all the patients of a practice to the area it lies within and viceversa - this is a case of spatial misalignment. Several approaches have been proposed to model spatially misaligned data and we review here the most relevant methods. 
	
	First of all, despite the fact that in a general sense this problem could be framed as change of support problem (COSP), given that the MSOAs making up the GPs' population are not guaranteed to be contiguous to each other, the classical \textit{block-averaging} approach \citep{Cressie1993} used for COSP becomes difficult to implement.
	A different approach is spatial point data modelling, by treating MSOA centroids and GP locations as observation sites. We would obtain two misaligned sets of locations and in that case spatial Gaussian processes modelling, such as the one presented in \citet{Ren2013}, could be a viable modelling strategy, since the observations for the two outcomes come from different sets of locations. However, the above mentioned problem of population location for the GPs would persist. Finally, an alternative approach is the \textit{non-nested block-level realignment} (\cite{Banerjee2014}, Section 7.3.2). In their application they overlay two non-nested grids, one made up of \textit{target zones} and the other of \textit{source zones}, obtaining a new one whose areas they refer to as \textit{atoms}. Their interest is in modelling a variable recorded on the target zones, by using covariates registered on the source zones. To achieve that they suggest a fully model based approach where the source zone covariates are modeled as a sum of latent variables defined on the atoms.  
	However, as GP populations can be scattered in non contiguous MSOAs, no unique areal subdivision of outer northern east London (ONEL) is available for prevalence data that could be used to pinpoint source zones. 
	In contrast, we propose to model diabetes prevalence at GP level via a weighted combination of covariates and random effects, determined using multiple membership, defined on a spatial framework determined by the MSOAs.

	The data framework used here comprises 130 GP practices in three Clinical Commissioning Groups (CCGs) in outer northern east London (ONEL).  MSOAs were included (in ONEL and beyond) if their residents were among the largest contributors to the selected GP populations. This procedure results in 95 MSOAs, 88 of which are actually in ONEL. With this data we aim to analyse prevalence and mortality risk for MSOAs using prevalence information for GP practices.
	
	To the best of the authors’ knowledge, there are no previous Bayesian studies of the spatial causal relationship between diabetes prevalence and mortality. One aim of the study is to examine the strength of this relationship (if any) at a small area scale. This relationship may be attenuated at individual patient level because diabetes as a disease can be followed by cause by death from many other conditions, thus making more difficult to list it as cause of death. Consequently, \citet{McEwen2006} report that of nearly 12,000 diabetic patients in the US, diabetes was recorded on a minority (39\%) of death certificates, and as the underlying cause of death for only 10\% of decedents with diabetes. There is also the issue of accurate recording: many studies report that diabetes is an underreported cause of death. \citet{Stokes2017} report that the proportion of deaths with diabetes assigned as the underlying cause of death (3.3–3.7\%) severely understates the contribution of diabetes to mortality in the United States.
	
	Therefore, the main contribution of this paper is methodological: the incorporation of the multiple membership approach into a spatial prior. Previous applications of multiple membership have been in crossed multilevel applications (e.g. educational careers for pupils within schools) with unstructured priors. Another significant contribution is computational, as to the best of our knowledge this is the first implementation of a bivariate CAR spatial prior into Rstan \parencite{Carpenter2017,Team2018}, a probabilistic programming platform which performs full Bayesian inference using Hamiltonian Monte Carlo (HMC). Previous Rstan implementations or CAR priors have been univariate \parencite{Morris2019,Joseph2016}. Lastly, there is novelty in the application of the methods to the specific disease studied, and in showing how spatial priors can be applied to distinct health outcomes (e.g. prevalence, mortality) where causal precedence is clear.
	
	This latter feature of our proposed modelling approach embeds causal information in the estimation of the spatial random effects. In our work causality arises from the necessary precedence of prevalence over mortality, therefore we embed this information in our model by using the GMCAR prior \parencite{Jin2005}. This prior allows to estimate spatial random effects using a conditional formulation, which in our case entails specifying spatial random effects for mortality conditional on the ones for prevalence. While others have proposed models that study the spatial causal structure to adjust for unmeasured confounding in a fully Bayesian framework \citep{Schnell2019}, this is beyond the scope of our work.
	
	The paper is organised as follows. We briefly describe the motivating dataset in Section \ref{sec:data}. In Section \ref{sec:CAR} we review conditional autoregressive priors for modelling bivariate area disease counts and in Section \ref{sec:mmb_areal} we incorporate the multiple membership approach into a spatial prior. We discuss the implementation of this approach in Stan in Section \ref{sec:impl}. A simulation study is included in Section \ref{sec:sim} and findings regarding the motivating dataset on diabetes prevalence and mortality are presented in Section \ref{sec:realdata}. 
	
	\subsection{Data structure} \label{sec:data}
	
	Data on diabetes mortality is available from Census units called Middle Layer Super Output Areas (MSOAs), while data on diabetes prevalence is available for GP disease registers. With this data we aim to analyse prevalence and mortality risk for MSOAs using prevalence information for GP practices. 
	
	The data framework comprises all 130 GP practices in three Clinical Commissioning Groups (CCGs) in outer northern east London (ONEL): Barking and Dagenham, Havering and Red-bridge.  MSOAs were included if their residents were among the largest contributors to the selected GPs populations up to the 90th percentile. This procedure results in 95 MSOAs, 88 of which are actually in ONEL. 
	
\begin{figure}[ht]
		\centering
		\includegraphics[scale=.04]{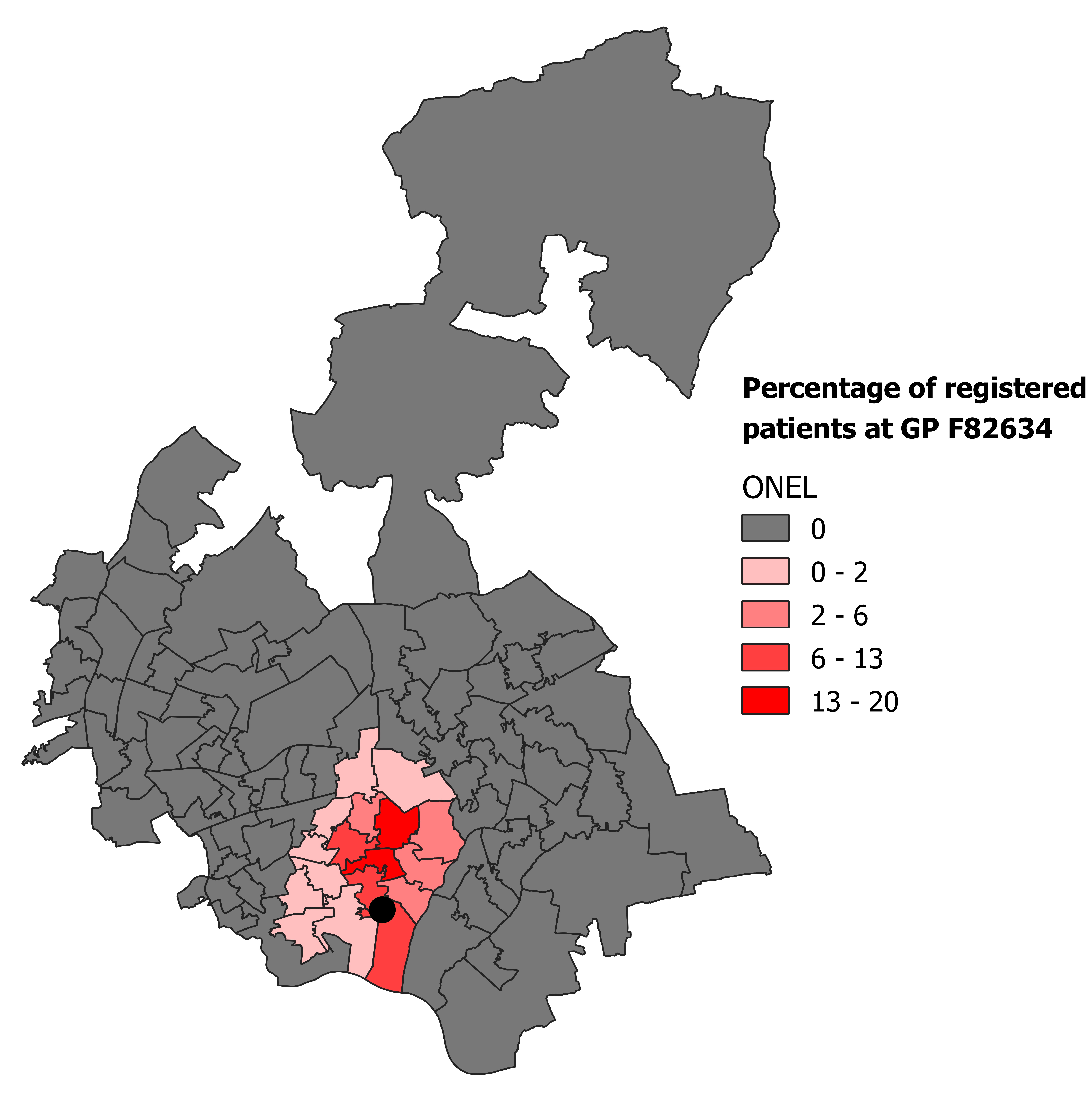}
		\caption{Example of contributing MSOAs populations for a specific GP practice, identified by a black dot on the map.
		}
		\label{fig:mm_ex}
	\end{figure}

	The misalignment must be modelled carefully as it is significant: on average 43.4 MSOAs contribute to the population of each GP (median 41 and range 14 to 80), and the average contribution of each MSOA to the population of a single GP is 2.3\% (median 0.19 \%) ranging
	between 0.0059\% to 96 \% (see Figure \ref{fig:mm_ex} as an example of a population catchment area for a GP practice).
	
	Mortality data by MSOA include death counts from 2013 to 2017. Consequently October 2015 was chosen as midpoint for retrieving prevalence data, and also for residential patient flow data for each GP; the latter provides a cross reference file linking MSOA and GP populations. In addition to that, two covariates recorded at the MSOA level have been included, namely the proportion of South Asian residents from the 2011 Census \footnote{Which include the following ethnicities: Indian, Pakistani and Bangladeshi, taken from Table KS201EW in the NOMIS database} and the 2015 Index of Multiple Deprivation (IMD). The former variable has been included due to the higher prevalence of diabetes among the South Asian population \parencite{Hall2010}, while the latter is a control variable for area socio-economic status.
	
	For the purpose of estimating areal prevalence and mortality, it is necessary to take into account population structure, therefore expected counts for each area are computed using national rates for mortality and prevalence, together with the number of residents by age. This gives 
	\begin{align} \label{eq:offsets}
		E_{ik} = \sum_{a=1}^{A}\omega_{ak}  N_{ik}^a
	\end{align}
	where $ i=1,...,n $ is the area, $ k = 1,...,K $ the outcome, $ a=1,\ldots,A $ is the population age group, $ \omega_{ak} $ are the national mortality and prevalence rates for diabetes broken down by age group together with $ N_{ik}^a $ the corresponding MSOA or GP populations \parencite{Blangiardo2015}. In the case of prevalence offset computation, index \textit{i} for MSOAs must be substituted by $ j =1,...,m $ indexing GPs instead.

	\section{Modelling and conditional autoregressive priors (CAR)} \label{sec:CAR}

	Generalised linear models (GLM) with spatially correlated random effects are commonly used in spatial data analysis. We consider a spatial GLM with Gaussian CAR random effects where a region is partitioned in \textit{n} non-overlapping areas. For each area $ i = 1, \ldots, n $ we can observe a random variable representing the outcome of interest $ Y_i $ and a set of \textit{p} non-random covariates $ \bm{x}^T_i = (x_{i1},\ldots, x_{ip})^T $. Define $ E_i $ as known offset values, and $ \rho_i $ as the relative disease risk \parencite{Blangiardo2015}. 
	
	The methods in this paper apply to all GLMs but, without loss of generality, in this paper we will model count data using the Negative Binomial distribution with mean $ \mu_i $ for each observation, a common overdispersion parameter $ \psi $ and a logarithmic link function. The Negative Binomial allows explicit modelling of overdispersion, when the variance in the outcome is larger than the average outcome \parencite{Coly2019}.
	
	We can then write the GLM with the CAR term $ \phi_i $, defined below, as:
	\begin{align}
		\label{eq:glm1}
		\begin{split}
			Y_{i} &\sim \mbox{NegBin}(E_i\rho_i, \psi) \\
			\log(\rho_i) &= \gamma + \bm{x}^T_i\bm{\beta} + \phi_i 
		\end{split}
	\end{align}
	To model explicitly the overdispersion, we set $ \mu_i = E_i\rho_i $ so it follows that
	\begin{align} \label{eq:negbin_den}
		f_Y(y_i; \mu_i, \psi) = \begin{pmatrix}
			y_i+\psi-1 \\ y_i
		\end{pmatrix} \left(\frac{\mu_i}{\mu_i+\psi}\right)^{y_i}\left(\frac{\psi}{\mu_i+\psi}\right)^\psi \quad \forall i = 1,\ldots,n
	\end{align}
	where $ y \in \mathbb{R}^+ $; $ E(Y_i) = \mu_i \in \mathbb{R}^+ $, $ \psi \in \mathbb{R}^+ $ and $ Var(Y_i)  = \mu_i + \mu_i^2 / \psi$.

	When considering areal data the set of areas on which data are recorded can either form a regular lattice or differ largely in both shape and size. In either case such data typically exhibit spatial autocorrelation, with observations from areal units close together tending to have similar values. Consider the case where no spatial misalignment is present by assuming a \textit{Gaussian Markov random Field} (GMRF) on the lattice \parencite{Rue2005}. Then it is possible to define $ \phi_i $ through the following full conditionals,
	\begin{align}
		\phi_i | \phi_{-i} \sim N(\alpha \sum_{j \in \partial i}b_{ij} \phi_j, \tau_i^{-1}) 
	\end{align}
	where $ b_{ij} $ is an element of matrix \textit{B} defined below,  red$ \tau_i^{-1} = (\tau/d_i)^{-1}  $ with $ \tau $ scaling the marginal precision, $ d_i $ is the number of neighbours for area \textit{i}, $\partial i$ is the set of all neighbours of \textit{i} and $ \alpha $, a propriety parameter, is usually interpreted as a smoothing parameter. 
	Thanks to Brook's Lemma \parencite{Rue2005} it is guaranteed that the joint distribution is proper, and it is given by
	\begin{align} \label{eq:joint_uni_car}
		\begin{split}
			\bm{\phi} \sim \mathcal{N}_n(\bm{0},& \left[\tau D(I - \alpha B)\right]^{-1})\\
			\text{where } \,\, & 0<\alpha<1
		\end{split}
	\end{align}
	where $ B = D^{-1}W$, $ D = diag(d_i) $,  and \textit{W} is the adjacency matrix such that $ (W)_{ii} = 0 $ and $ (W)_{ij} = 1 $ if \textit{i} and \textit{j} are neighbours. This setup can give rise to different prior choices that require different conditions to achieve properness \parencite{Congdona}, for instance see \citet{Stern1999} for a different choice of spatial weights for matrix \textit{B} based on expected counts.
	
	Building from \citet{Mardia1988} we can specify a \textit{K}-dimensional multivariate spatial model. However, general conditions for  positive definiteness and invertibility for the covariance matrix are difficult to assess. Therefore, we use the following formulation by \citet{Gelfand2003}
	\begin{align} \label{eq:MCAR_p}
		\bm{v} \sim\mathcal{N}_{nK}\left(\bm{0}, \left[ \Lambda \otimes (D-\alpha W) \right]\right)^{-1}
	\end{align}
	where $ \Lambda $ is a $ K\times K $ matrix that controls the variance and covariance between outcomes in the same area, and is therefore required to be symmetric positive definite. This model will be denoted as $ \text{MCAR}(\alpha, \Lambda) $.
	
	An MCAR prior for spatial random effects allows estimates of relative risks to borrow strength from each other. This is a reasonable assumption for our motivating example, as diabetes prevalence and mortality are correlated. The MCAR on spatial effects can capture such relationship and improve the estimate for both outcomes.
	
	When considering prevalence and mortality for the same disease, given the necessary precedence of prevalence on mortality, it is possible to draw a causal link between the two outcomes. In order to embed that information in the model we can use the GMCAR \parencite{Jin2005}, a specification of the previously mentioned MCAR for the bivariate case, that is based on estimating the spatial effects for one outcome conditionally on the other, as follows.
	\begin{align} 
		\label{eq:gmcar}
		\begin{split}
			\bm{\phi_1} | \bm{\phi_2} & \sim \mathcal{N}_n
			((\eta_0I_n + W\eta_1)\bm{\phi_2}, \left[\tau_1(D-\alpha_1W)\right]^{-1})\\
			\bm{\phi_2} & \sim \mathcal{N}_n(\bm{0},
			\left[\tau_2(D-\alpha_2W)\right]^{-1}) \\
			\text{with } & \quad |\alpha_1|<1 \,\, \text{and} \,\, |\alpha_2|<1	\end{split}
	\end{align}
	In our context, $\bm{\phi}_1|\bm{\phi}_2$ would be the spatial random effect for mortality given the spatial structure of prevalence, while $\bm{\phi}_2$ would be the spatial random effect for prevalence. It should be noted that this model requires a conditional specification of said random effects. Therefore the resulting estimated parameters will depend on the order chosen for the outcomes.
	
	In (7) the parameter $\alpha_1$ is the smoothing parameter associated with the conditional distribution of $\bm{\phi}_1|\bm{\phi}_2$, $\alpha_2$ is similarly defined for the marginal distribution of $\bm{\phi}_2$, and $\tau_1$ and $\tau_2$ scale the precision of $\bm{\phi}_1|\bm{\phi}_2$ and $\bm{\phi}_2$, respectively. As the conditional expected value is linear in $ \eta_0 $ and $ \eta_1 $, these two parameters reflect the linear relationship between the two diseases risks. The parameter $ \eta_0 $  represents area specific (own area) spatial dependence, while $\eta_1 $ refers to the neighbour effect between the two outcomes. We will refer to this model as $\mbox{GMCAR}(\alpha_1,\alpha_2,\eta_0,\eta_1, \tau_1,\tau_2)$.
	A particularly interesting feature of model (\ref{eq:gmcar}) is that for the bivariate case by setting $ \eta_1 =0 $ and $ \alpha_1 = \alpha_2 $ we obtain the $\text{MCAR}(\alpha,\Lambda) $ of equation (\ref{eq:MCAR_p}). Additionally, referring to standard conditional Normal distribution theory (\cite{Rue2005}; Section 2.1.7) we can more explicitly outline the functional relationships between the elements of matrix $\Lambda$ of equation (\ref{eq:MCAR_p}) and the GMCAR parameters since: $\tau_1 = \Lambda_{11}$, $\tau_2 = \Lambda_{22} - \Lambda_{12}^2/\Lambda_{11}$ and $\eta_0 = - \Lambda_{12}/\Lambda_{11}$.

	Section 3 in \citet{Jin2005} shows how to generalise the GMCAR model to K outcomeswithout removing the order dependence, by defining conditional distributions sequentially. In the present context, one might (for example) take obesity as preceding both diabetes prevalence and mortality. For $K = 3$ the joint density of the random effects can be written as:
	\begin{align}
		p(\bm{\phi}) = p(\bm{\phi}_1 \mid  \bm{\phi}_2, \bm{\phi}_3) p(\bm{\phi_2} \mid \bm{\phi}_3) p(\bm{\phi}_3)
	\end{align}
	with $ \bm{\phi}^T = (\bm{\phi}_1^T, \bm{\phi}_2^T, \bm{\phi}_3^T) $ and properness trivially satisfied.
	
	Finally, it should be mentioned the GMCAR can be further generalised to a K-variate order-free model via a co-regionalisation approach as shown in \citet{Jin2007}
	
	\section{Multiple membership for areal misalignment} \label{sec:mmb_areal}
	
	In this paper we leverage the MM principle \parencite{Browne2001} by building memberships with averages of areal relative risks. Let $\bm{Y}_k = (Y_{1k} , ... , Y_{m_k k})$ be the vector of observed counts for outcome $k = 1, ..., K$ with $m_k$ memberships. We will define the relative risk of each membership of $\bm{Y}_k$ by a weighted average of areal relative risks as explained in Section \ref{sec:CAR}. Indicating areas with $i = 1, ..., n$, we can specify a set of weights for each membership $ j = 1,..., m_k $ as:
	\begin{align}	\label{eq:mm_weights}
		h_{i \mid j,k} \, \, s.t. \,\, \sum_{i = 1}^{n} h_{i|j,k} = 1
	\end{align}
	where obviously $h_{i \mid j,k} = 0$ for the areas \textit{i} that do not contribute to membership \textit{j}. 
	
	Referring back to Equation \ref{eq:glm1}, where we introduced areal relative risk for area \textit{i}, the new specification for membership \textit{j} and outcome \textit{k} will be:
	\begin{align} \label{eq:mm_rr}
		\begin{split}
			Y_{jk} &\sim NegBin(E_{jk} \rho_{jk}, \psi_k) \\
			\log(\rho_{jk}) &= \sum_{i = 1}^{n} h_{i|j,k} (\gamma_k + \bm{x}^T_{ik}\bm{\beta}_k + \phi_{ik})
		\end{split}
	\end{align}
	where $E_{jk}$ are as in Equation \ref{eq:offsets}. 
	
	Given appropriate weights and memberships, this setup can be extended to an indefinite number of outcomes. However, in the present analysis we limit ourselves to two outcomes, and we apply the scheme to deal with the misalignment arising from the discrepancy in the way mortality and prevalence data are collected. Firstly, patients from different and non neighbouring MSOAs can register at the same GP (i.e. the mapping from MSOAs to GPs is not injective) and, secondly, individuals from the same MSOA can register at different GPs. 
	
	We obtain practice specific weights based on the relative contribution of each MSOA (in terms of population affiliated) to each practice.  In the dataset we include non-zero weights only for the MSOAs with the largest contributions, namely those constituting up to 90\% coverage of the GP practice population.
	
	For each practice different MSOAs contribute with a different weight, so the idea is that the prevalence relative risk for each GP is then computed as an average of MSOA level risks.  We can safely drop the second subscript \textit{k} for $ h_{i|j,k} $ from Equation \ref{eq:mm_weights}, since we have only one MM outcome. Specifying the model from Equations \ref{eq:glm1} and \ref{eq:mm_rr}
	\begin{eqnarray}\label{eq:full}
		MSOAs&:& i = 1,\ldots,n \\
		GPs &:& j = 1,\ldots,m\\
		y_{i1} &\sim& NegBin(E_{i1}\rho_{i1}, \psi_1)  \\ y_{j2} &\sim& NegBin(E_{j2}\rho_{j2}, \psi_2) \\
		\log(\rho_{i1}) &= & \gamma_1 + x_i\bm{\beta_1} + \phi_{i1}\\	
		\log(\rho_{j2})  &=& \sum_{i = 1}^{n}h_{i|j}(\gamma_2 + x_i\bm{\beta_2} + \phi_{i2}) 
	\end{eqnarray}
	where the spatial random effects are modelled as either 
	\begin{eqnarray} \label{eq:GMCAR_name}
		\bm{\phi}_1,\bm{\phi}_2 &\sim& \mbox{GMCAR}(\alpha_1,\alpha_2,\eta_0,\eta_1, \tau_1,\tau_2)
	\end{eqnarray}
	or
	\begin{eqnarray} \label{eq:MCAR_name}
		\bm{\phi}_1,\bm{\phi}_2 &\sim& \mbox{MCAR}(\alpha,\Lambda).
	\end{eqnarray}
	We remark that equations \ref{eq:GMCAR_name} and \ref{eq:MCAR_name} refer to spatially structured models on a lattice as explained in Section \ref{sec:CAR}, and not considering the MM weighting.
    Priors are defined for parameters $\bm{\beta}, \alpha_1, \alpha_2, \gamma_1, \gamma_2, \eta_0, \eta_1, \tau_1, \tau_2, \psi_1, \psi_2$, covariates coefficients $ \bm{\beta}_{j} = (\beta_{j1}, \ldots, \beta_{jP}) $ with $ j = 1,2 $ indexing the outcome (1 for mortality and 2 for prevalence).
	When implementing the MM aspect of the model we indicate with $ h_{i|j} $ each area's weight for a specific membership; we refer to this as either GMCAR-MM or MCAR-MM, depending on whether the GMCAR or MCAR prior is used on the underlying areal spatial structure. 
	
	Since we are mostly interested in relative risk by area rather than membership, we introduce the following notational shortcut to denote the estimated areal component for outcomes observed by membership. We then refer to Equation \ref{eq:mm_rr} as:
	\begin{align} \label{eq:rr_membToArea}
		\sum_{i = 1}^{n} h_{i|j}(\gamma_k + \bm{x}_{ik}^T\bm{\beta}_k + \phi_{ik}) = \sum_{i = 1}^{n}h_{i|j}\zeta_{ik}
	\end{align}
	so that $ \zeta_{ik} = \gamma_k + \bm{x}^T_{ik}\bm{\beta}_k + \phi_{ik} $ indicates the relative risk for area \textit{i} implicitly defined by multiple membership.
	
	\subsection{Implementation}\label{sec:impl}
	
	To the best of our knowledge, this is the first implementation of multivariate CAR models (GMCAR and MCAR) in RStan \parencite{Carpenter2017}. The code is available on github at \url{https://github.com/silvialiverani/GMCARMM}.
	
	Since the MCAR prior can be viewed as a special case of the GMCAR (see Section \ref{sec:CAR}), the latter was coded first, while the implementation of the former was achieved by adding the necessary constraints, i.e. $ \eta_1 = 0 $ and $ \alpha_1=\alpha_2 $.
	
	We implemented a QR parameterisation (Section 1.2 of \citet{Team2018}) for the estimation of the parameters $\bm{\beta}$, which exploits the fact that any design matrix $X$ can be decomposed using the thin QR decomposition into an orthogonal matrix $Q$ and an upper-triangular matrix $R$, i.e. $X=QR$. This parameterisation performs much better in practice, often both in terms of elapsed time and in terms of effective sample size, as it makes it easier for the Monte Carlo algorithm to move in the space when an informative prior on the location of $\bm{\beta}$ is not available. Moreover, before obtaining the QR decomposition we have also implemented a min-max normalisation of the covariates.
	
	Sampling for both covariance matrices of the CAR component was coded using unit precision, i.e. setting $ \tau_1 = \tau_2 = 1 $ in equation (\ref{eq:gmcar}), and rescaling the $ \bm{\phi}_1 $ and  $ \bm{\phi}_2 $. Additionally, since calculating the determinant for those covariance matrices would be extremely computationally expensive, \citet{Joseph2016} implemented an eigenvalue decomposition in Stan, as suggested by \citet{Jin2005}, reducing the computational burden significantly.
	
	We used flat priors for the parameters $\bm{\beta}$, while for the other parameters we used independent, proper, weakly informative prior distributions, as recommended by \citet{Gelman2013a}.

	Convergence was assessed with the $\hat{R}$ diagnostic \parencite{Gelman1992}, and accordingly satisfactory convergence is achieved for values below 1.01 \parencite{Vehtari2019}. 
	
	\subsection{Risk clustering}
	\label{sec:risk_clustering}
	Following \citet{Congdon2019}  we will assess spatial clustering by comparing the average relative risk over the HMC iterations for each area with both their neighbouring areas, or locality, and the other outcome, thus measuring bivariate clustering. 
	
	An adaptation of LISA (Local Indicator of Spatial Association), proposed by \citet{Laohasiriwong2017}, will be used: firstly, we compute the probability of a certain area having an excess relative risk with high probability. This is achieved by computing the percentage of the sampled relative risks in the HMC chains over a certain threshold (usually taken as 1). Therefore, for area \textit{i}, outcome \textit{k} and probability threshold $ T_R $ we have, using the $ s = 1,..., B$ HMC samples:
	\begin{align} \label{eq:rr_prob}
		P_H(i,k) = \frac{1}{B}\sum_{s=1}^{B} \mathbb{I}\left(\rho_{ik}^{(s)} > T_R \right)
	\end{align}
	where $ \I(\cdot) $ is the indicator function. The use of Monte Carlo iterations allows to account for uncertainty in the risk assessment.
	
	Secondly, once these quantities are obtained, corresponding probabilities for each of the localities are computed as follows. Substituting relative risk for area \textit{i} and iteration \textit{s} with the average relative risk of locality ($ \rho_{\sim ik}^{(s)} $), Equation (\ref{eq:rr_prob}) becomes:
	\begin{align} \label{eq:rr_prob_area}
		P_H(\sim i,k) = \frac{1}{B}\sum_{s=1}^{B} \mathbb{I}\left(\rho_{\sim ik}^{(s)} > T_R \right)
	\end{align}
	where we define $ \rho_{\sim ik}^{(s)} $, using adjacency matrix \textit{W} from Section \ref{sec:CAR}:
	\begin{align} \label{eq:rr_prob_neigh}
		\rho_{\sim ik}^{(s)} = \frac{1}{d_i} \sum_{j=1}^{d_i} (W)_{ij} \rho_{jk}^{(s)}
	\end{align}
	and $ d_i = \sum_{j=1}^{n} (W)_{ij}  $ is the number of areas in the locality \textit{i}.
	
	Finally, with both $ P_H(i,k) $ and $ P_H(\sim i,k) $ for each area, we can evaluate if area \textit{i} is the centre of a high or low risk cluster by comparing if both $ P_H $'s are above threshold $ T_P $ (usually 0.9 or 0.95).
	With this type of comparison, we are able to classify each area in one of 4 different categories based on either $ P_H(i,k) $ or $ P_H(\sim i,k) $ being above $ T_P $. These four categories are High-High (HH), High-Low (HL), Low-High (LH) and Low-Low (LL). 
	
	When an area's risk is compared with the respective locality risk for each outcome separately, we can obtain a bivariate classification of areas. Since we can obtain 4 categories for each outcome, for bivariate clustering we will have 16, as shown in Table \ref{tab:bivar_clust}. However, in order to simplify the interpretation of a map (see Figure \ref{fig:clust_biv}), only four categories will be represented by focusing on within area comparisons e.g. for an area with high relative risk for mortality and low risk for prevalence we will write \textit{M:H-P:L}.
	
	\subsection{Model comparison} \label{sec:mod_comp}
	To choose among competing models, we used the deviance information criterion, or DIC \parencite{Spiegelhalter2002}. This criterion is based on the posterior distribution of the deviance statistics.
	For simplicity of notation, we will refer to the parameters of the Negative Binomial with the vector $ \bm{\theta} = (\mu_1, \ldots, \mu_n, \psi) $ so we can write its saturated deviance as:
	\begin{align}
		D(\bm{\theta}) = 2\sum_{i=1}^{n}y_i \log\left(\frac{y_i}{\mu_i}\right) - (y_i+\psi)\log\left(\frac{1+y_i/\psi}{1+\mu_i/\psi}\right)
	\end{align}
	We compute both the average posterior deviance $ \overline{D(\bm{\theta})} $ and the deviance of averages of parameters' posterior  $ D(\bar{\bm{\theta}}) $. For the former, when all aspects of the model are assumed true,  the approximation $ \overline{D(\bm{\theta})}\approx n $ holds \parencite{Spiegelhalter2002}, where \textit{n} is the number of observations. Consequently, this can also be used as a measure of fit. 	Computing the posterior deviances allows to easily compute also the effective number of parameters, $ p_D = \overline{D(\bm{\theta})} - D(\bm{\theta}) $.

	Moreover, lack of fit of the data with respect to the posterior predictive distribution can be measured by the tail-area probability (TAP), or $p$-value, of the test quantity \parencite{Gelman2013a}. For each observation $ i = 1,\ldots,n $ and each outcome we can estimate:
	\begin{align}
		\label{eq:ppp}
		p_i = P(y^{rep}_i < y_i |y) + \frac{1}{2}P(y^{rep}_i = y_i |y)
	\end{align}
	where the $ y^{rep}_i $ are sampled at each HMC iteration from the likelihood conditional on the current sampled parameters. The probabilities are estimated by the proportion of times the events $y^{rep}_i < y_i$ or $ y^{rep}_i = y_i $ occur throughout the chain. In case of an adequate model the $ p_i $'s are expected to concentrate around 0.5 and away from the tails of the distribution, so by evaluating the tails of the estimated TAP's we can see how frequently the replicated values over (or under) predict the observed ones. Specifically  we consider the proportion of TAP's that higher than 0.95 and lower than 0.05, as well as higher than 0.90 and lower than 0.1. 
	
	We also evaluate the approximate leave-one-out cross-validation (LOO) \parencite{Vehtari2017}, using \textit{elpd}, the expected log pointwise predictive density for a new dataset, which is a sum of individual log predictive densities. We compare models (say A and B) by computing mean and standard error of the difference in \textit{elpd}:
	\begin{align} \label{eq:std_diff_elpd}
		se(\widehat{elpd}^A_{loo} - \widehat{elpd}^B_{loo}) = \sqrt{n V_{i=1}^n(\widehat{elpd}^A_{loo,i} - \widehat{elpd}^B_{loo,i}) }
	\end{align}
	where $ V_{i=1}^n(x)$ is the sample variance, i.e. $ V_{i=1}^n(x) = \frac{1}{n-1} \sum_{i=1}^{n} (x_i - \bar{x}) $.

	\section{Simulation study}\label{sec:sim}
	
	To validate our implementation we coded the GMCAR-MM and MCAR-MM models with multiple membership in BUGS and compared it with the Rstan runs. The BUGS (Bayesian inference Using Gibbs Sampling) project \parencite{Lunn2000} is concerned with flexible software for the Bayesian analysis of complex statistical models using Markov chain Monte Carlo (MCMC) methods. Although BUGS is a very popular software for Bayesian hierarchical modelling, it is important to note that it is no longer maintained. We compared our Rstan implementation of the models above to an implementation in R2OpenBUGS \parencite{Sturtz2005} (referred to as BUGS henceforth). 
	
	We generated four datasets: one with GMCAR-MM to simulate spatially varying effects with covariates, one with GMCAR-MM to simulate spatially varying effects without covariates, one with MCAR-MM to simulate spatially varying effects with covariates, one with MCAR-MM to simulate spatially varying effects without covariates.
	We chose realistic parameter values in our simulation, in line with the posterior means of the results on real data in Section \ref{sec:realdata}, as well as the true ONEL spatial lattice of 95 MSOAs, the offsets and true covariates mentioned in Section \ref{sec:data} to generate new observations for mortality and prevalence. Keeping these parameters fixed, we simulated each dataset 100 times type and estimated the corresponding model using both BUGS and RStan.
	
	With regard to the likelihood direct implementation of the Negative Binomial, as parametrised in equation (\ref{eq:negbin_den}), is not available in BUGS; therefore we implemented it using the equivalent formulation via a Poisson-Gamma mixture \parencite{Hilbe2011}. As prior distributions we used Half-Normals for the precision parameters $ \tau_j $, Gammas for the Negative Binomial overdispersions $\psi_j$ and Normal distributions for the intercepts $\gamma_j$ and linking parameters $\eta_0$ and $\eta_1$. Since the QR parametrization is not available in BUGS, with covariates coefficients $ \bm{\beta} $ we used very vague Normal priors.

	Tables \ref{tab:stan_bugs_nocov}, \ref{tab:stan_bugs_cov_gm} and \ref{tab:stan_bugs_cov_mv} show summaries of the distribution of the posterior means obtained across the 400 simulated datasets (100 for each model). Each posterior was obtained by running 4 chains each with 50\% warm-up and 6000 iterations for BUGS, while RStan required only 3000 iteration per chain to achieve convergence. Given the different number of iterations employed for the two packages, computational times are lower for RStan since each run took on average 4'10'' to complete against BUGS's 9'20''.  Simulations were run on Queen Mary's Apocrita HPC facility, supported by QMUL Research-IT \parencite{king_thomas_2017_438045}.
	
	The results show that RStan reports better convergence as the percentage of $ \hat{R} $ below 1.1 or even 1.01 are significantly lower than in the BUGS case. The 95\% posterior credible intervals generally contain the true values, both programs report similar results in this regard. 
	
	\begin{table}
		\centering
		\caption{Summary descriptive statistics of posterior means for the parameters of interest in the 100 simulations (models \textit{without} covariates). BUGS (grey lines) and Rstan (white lines). Results from running 4 chains, 6000 iterations for BUGS, 3000 for RStan and 50\% warm-up for both. $\%\hat{R}>t$ indicates the percentage of $ \hat{R} $ over threshold \textit{t} in the 100 simulations}
		\label{tab:stan_bugs_nocov}
		\begin{tabular}{lr} 
			\rowcolors{2}{Lightgray}{white}
			\begin{tabular}{Arrrrrrr}
				\toprule
				\multicolumn{8}{c}{\textbf{GMCAR-MM}} \\
				& True & mean & sd & 2.5\% & 97.5\% & $\%\hat{R}>1.01$ & $\%\hat{R}>1.1$ \\ 
				\midrule
				$\alpha_1$ & 0.40 & 0.50 & 0.03 & 0.45 & 0.56 & 0.00 & 0.00 \\ 
				$\alpha_1$ & 0.40 & 0.50 & 0.04 & 0.45 & 0.57 & 3.00 & 1.00 \\ 
				$\alpha_2$ & 0.90 & 0.65 & 0.15 & 0.39 & 0.91 & 44.00 & 0.00 \\ 
				$\alpha_2$ & 0.90 & 0.66 & 0.14 & 0.38 & 0.91 & 2.00 & 0.00 \\ 
				$\tau_1$ & 6.00 & 6.10 & 0.63 & 4.93 & 7.14 & 76.00 & 0.00 \\ 
				$\tau_1$ & 6.00 & 6.10 & 0.67 & 4.79 & 7.18 & 0.00 & 0.00 \\ 
				$\tau_2$ & 4.00 & 5.05 & 1.42 & 2.57 & 7.37 & 100.00 & 12.00 \\ 
				$\tau_2$ & 4.00 & 4.87 & 1.58 & 2.16 & 7.48 & 2.00 & 1.00 \\ 
				$\eta_0$ & 0.30 & 0.50 & 0.35 & -0.19 & 1.16 & 100.00 & 26.00 \\ 
				$\eta_0$ & 0.30 & 0.48 & 0.33 & -0.11 & 1.17 & 2.00 & 1.00 \\ 
				$\eta_1$ & 0.50 & 0.57 & 0.17 & 0.32 & 0.90 & 100.00 & 22.00 \\ 
				$\eta_1$ & 0.50 & 0.57 & 0.17 & 0.30 & 0.86 & 3.00 & 1.00 \\ 
				$\psi_1$ & 20.00 & 21.19 & 3.61 & 14.64 & 27.50 & 39.00 & 0.00 \\ 
				$\psi_1$ & 20.00 & 21.20 & 3.87 & 12.91 & 28.28 & 0.00 & 0.00 \\ 
				$\psi_2$ & 10.00 & 10.00 & 1.11 & 7.96 & 12.21 & 7.00 & 0.00 \\ 
				$\psi_2$ & 10.00 & 10.34 & 1.40 & 8.52 & 13.22 & 1.00 & 1.00 \\ 
				$\gamma_1$ & 0.50 & 0.50 & 0.18 & 0.14 & 0.81 & 81.00 & 0.00 \\ 
				$\gamma_1$ & 0.50 & 0.49 & 0.19 & 0.12 & 0.80 & 7.00 & 1.00 \\ 
				$\gamma_2$ & 1.30 & 1.30 & 0.06 & 1.19 & 1.41 & 89.00 & 2.00 \\ 
				$\gamma_2$ & 1.30 & 1.31 & 0.07 & 1.19 & 1.42 & 4.00 & 0.00 \\ 
				\bottomrule
			\end{tabular}
		\end{tabular}
		\quad
		\rowcolors{2}{Lightgray}{white}
		\begin{tabular}{Arrrrrrr} 
			\toprule
			\multicolumn{8}{c}{\textbf{MCAR-MM}} \\
			\toprule
			& True & mean & sd & 2.5\% & 97.5\% & $\%\hat{R}>1.01$ & $\%\hat{R}>1.1$ \\ 
			\midrule
			$\alpha$ & 0.40 & 0.43 & 0.09 & 0.30 & 0.66 & 11.00 & 0.00 \\ 
			$\alpha$ & 0.40 & 0.42 & 0.08 & 0.31 & 0.56 & 0.00 & 0.00 \\ 
			$\tau_1$ & 6.00 & 6.15 & 0.71 & 4.52 & 7.37 & 74.00 & 0.00 \\ 
			$\tau_1$ & 6.00 & 6.09 & 0.73 & 4.85 & 7.59 & 0.00 & 0.00 \\ 
			$\tau_2$ & 4.00 & 5.50 & 1.23 & 3.11 & 7.42 & 86.00 & 2.00 \\ 
			$\tau_2$ & 4.00 & 5.70 & 1.41 & 2.84 & 7.83 & 0.00 & 0.00 \\ 
			$\eta_0$ & 0.30 & 0.20 & 0.37 & -0.49 & 0.87 & 99.00 & 12.00 \\ 
			$\eta_0$ & 0.30 & 0.16 & 0.33 & -0.54 & 0.73 & 0.00 & 0.00 \\ 
			$\psi_1$ & 20.00 & 20.92 & 3.94 & 13.54 & 28.20 & 21.00 & 0.00 \\ 
			$\psi_1$ & 20.00 & 20.49 & 3.96 & 12.01 & 27.82 & 0.00 & 0.00 \\ 
			$\psi_2$ & 10.00 & 10.05 & 1.29 & 8.25 & 13.34 & 3.00 & 0.00 \\ 
			$\psi_2$ & 10.00 & 9.99 & 1.44 & 7.70 & 13.19 & 0.00 & 0.00 \\ 
			$\gamma_1$ & 0.50 & 0.49 & 0.06 & 0.38 & 0.59 & 61.00 & 0.00 \\ 
			$\gamma_1$ & 0.50 & 0.49 & 0.06 & 0.37 & 0.58 & 0.00 & 0.00 \\ 
			$\gamma_2$ & 1.30 & 1.30 & 0.04 & 1.24 & 1.37 & 1.00 & 0.00 \\ 
			$\gamma_2$ & 1.30 & 1.30 & 0.04 & 1.23 & 1.38 & 0.00 & 0.00 \\ 
			\hline
			\bottomrule
		\end{tabular}
	\end{table}

	\begin{table}
		\centering
		\caption{Summary descriptive statistics of posterior means for the parameters of interest in the 100 simulations (GMCAR \textit{with} covariates). BUGS (grey lines) and Rstan (white lines). Results from running 4 chains, 6000 iterations for BUGS, 3000 for RStan and 50\% warm-up for both. $\%\hat{R}>t$ indicates the percentage of $ \hat{R} $ over threshold \textit{t} in the 100 simulations}
		\label{tab:stan_bugs_cov_gm}
		\rowcolors{2}{Lightgray}{white}
		\begin{tabular}{Arrrrrrr} 
			\toprule
			\multicolumn{8}{c}{\textbf{GMCAR-MM with Covariates}} \\
			\midrule
			& True & mean & sd & 2.5\% & 97.5\% & \%$\hat{R}>1.01$ & \%$\hat{R}>1.1$ \\ 
			\hline
			$\alpha_1$ & 0.40 & 0.53 & 0.07 & 0.44 & 0.72 & 9.00 & 0.00 \\ 
			$\alpha_1$ & 0.40 & 0.50 & 0.04 & 0.44 & 0.58 & 1.00 & 0.00 \\ 
			$\alpha_2$ & 0.20 & 0.48 & 0.08 & 0.37 & 0.69 & 17.00 & 0.00 \\ 
			$\alpha_2$ & 0.20 & 0.44 & 0.07 & 0.36 & 0.63 & 1.00 & 0.00 \\ 
			$\tau_1$ & 6.00 & 5.19 & 0.73 & 3.74 & 6.55 & 86.00 & 1.00 \\ 
			$\tau_1$ & 6.00 & 5.63 & 0.66 & 4.20 & 6.83 & 0.00 & 0.00 \\ 
			$\tau_2$ & 4.00 & 5.80 & 1.19 & 3.24 & 7.39 & 95.00 & 5.00 \\ 
			$\tau_2$ & 4.00 & 6.24 & 1.37 & 3.09 & 8.48 & 0.00 & 0.00 \\ 
			$\eta_0$ & 0.30 & 0.59 & 0.54 & -0.32 & 1.61 & 99.00 & 42.00 \\ 
			$\eta_0$ & 0.30 & 0.34 & 0.48 & -0.55 & 1.19 & 3.00 & 0.00 \\ 
			$\eta_1$ & 0.50 & 0.04 & 0.10 & -0.16 & 0.19 & 98.00 & 7.00 \\ 
			$\eta_1$ & 0.50 & 0.35 & 0.24 & -0.18 & 0.75 & 5.00 & 0.00 \\ 
			$\psi_1$ & 20.00 & 16.84 & 3.57 & 9.85 & 24.25 & 56.00 & 0.00 \\ 
			$\psi_1$ & 20.00 & 18.52 & 3.33 & 12.02 & 24.04 & 0.00 & 0.00 \\ 
			$\psi_2$ & 10.00 & 10.21 & 1.24 & 8.20 & 13.15 & 5.00 & 0.00 \\ 
			$\psi_2$ & 10.00 & 9.85 & 1.21 & 7.44 & 12.15 & 0.00 & 0.00 \\ 
			$\gamma_1$ & -0.30 & -0.34 & 0.22 & -0.82 & 0.00 & 97.00 & 1.00 \\ 
			$\gamma_1$ & -0.30 & -0.32 & 0.20 & -0.69 & 0.03 & 1.00 & 0.00 \\ 
			$\gamma_2$ & -0.50 & -0.50 & 0.10 & -0.68 & -0.34 & 85.00 & 1.00 \\ 
			$\gamma_2$ & -0.50 & -0.50 & 0.09 & -0.66 & -0.33 & 1.00 & 0.00 \\ 
			$\beta_{11}$ & 0.30 & 0.31 & 0.30 & -0.23 & 0.87 & 85.00 & 0.00 \\ 
			$\beta_{11}$ & 0.30 & 0.28 & 0.32 & -0.29 & 0.87 & 0.00 & 0.00 \\ 
			$\beta_{12}$ & 0.50 & 0.55 & 0.34 & -0.00 & 1.27 & 100.00 & 2.00 \\ 
			$\beta_{12}$ & 0.50 & 0.47 & 0.32 & -0.01 & 1.07 & 0.00 & 0.00 \\ 
			$\beta_{21}$ & 1.00 & 1.19 & 0.15 & 0.91 & 1.43 & 32.00 & 0.00 \\ 
			$\beta_{21}$ & 1.00 & 1.20 & 0.13 & 0.94 & 1.43 & 1.00 & 0.00 \\ 
			$\beta_{22}$ & 1.00 & 1.00 & 0.17 & 0.70 & 1.32 & 84.00 & 2.00 \\ 
			$\beta_{22}$ & 1.00 & 0.99 & 0.15 & 0.70 & 1.30 & 0.00 & 0.00 \\
			\bottomrule
		\end{tabular}
	\end{table}
	
	\begin{table}
		\centering
		\caption{Summary descriptive statistics of posterior means for the parameters of interest in the 100 simulations (MCAR \textit{with} covariates). BUGS (grey lines) and Rstan (white lines). Results from running 4 chains, 6000 iterations for BUGS, 3000 for RStan and 50\% warm-up for both. $\%\hat{R}>t$ indicates the percentage of $ \hat{R} $ over threshold \textit{t} in the 100 simulations}
		\label{tab:stan_bugs_cov_mv}
		\rowcolors{2}{Lightgray}{white}
		\begin{tabular}{Arrrrrrr} 
			\toprule
			\multicolumn{8}{c}{\textbf{MCAR-MM with Covariates}} \\
			& True & mean & sd & 2.5\% & 97.5\% & $\%\hat{R}>1.01$ & $\%\hat{R}>1.1$ \\ 
			\midrule
			$\alpha$ & 0.40 & 0.45 & 0.07 & 0.34 & 0.61 & 16.00 & 0.00 \\ 
			$\alpha$ & 0.40 & 0.43 & 0.06 & 0.35 & 0.57 & 0.00 & 0.00 \\ 
			$\tau_1$ & 6.00 & 5.82 & 0.71 & 4.58 & 7.06 & 88.00 & 0.00 \\ 
			$\tau_1$ & 6.00 & 5.90 & 0.60 & 4.76 & 7.03 & 0.00 & 0.00 \\ 
			$\tau_2$ & 4.00 & 5.62 & 1.30 & 2.98 & 7.60 & 87.00 & 0.00 \\ 
			$\tau_2$ & 4.00 & 5.62 & 1.25 & 3.05 & 7.49 & 0.00 & 0.00 \\ 
			$\eta_0$ & 0.30 & 0.18 & 0.43 & -0.63 & 0.93 & 100.00 & 21.00 \\ 
			$\eta_0$ & 0.30 & 0.12 & 0.39 & -0.59 & 0.91 & 0.00 & 0.00 \\ 
			$\psi_1$ & 20.00 & 19.22 & 3.78 & 12.43 & 25.91 & 22.00 & 0.00 \\ 
			$\psi_1$ & 20.00 & 19.53 & 3.34 & 13.74 & 25.96 & 0.00 & 0.00 \\ 
			$\psi_2$ & 10.00 & 10.15 & 1.43 & 7.75 & 13.06 & 2.00 & 0.00 \\ 
			$\psi_2$ & 10.00 & 10.18 & 1.49 & 7.74 & 13.32 & 0.00 & 0.00 \\ 
			$\gamma_1$ & -0.30 & -0.33 & 0.16 & -0.65 & -0.03 & 95.00 & 6.00 \\ 
			$\gamma_1$ & -0.30 & -0.34 & 0.15 & -0.63 & -0.05 & 0.00 & 0.00 \\ 
			$\gamma_2$ & -0.50 & -0.51 & 0.10 & -0.70 & -0.34 & 55.00 & 0.00 \\ 
			$\gamma_2$ & -0.50 & -0.51 & 0.11 & -0.69 & -0.34 & 0.00 & 0.00 \\ 
			$\beta_{11}$ & 0.30 & 0.29 & 0.23 & -0.16 & 0.69 & 85.00 & 0.00 \\ 
			$\beta_{11}$ & 0.30 & 0.31 & 0.23 & -0.12 & 0.69 & 0.00 & 0.00 \\ 
			$\beta_{12}$ & 0.50 & 0.49 & 0.29 & -0.06 & 1.00 & 96.00 & 4.00 \\ 
			$\beta_{12}$ & 0.50 & 0.51 & 0.26 & 0.05 & 0.97 & 0.00 & 0.00 \\ 
			$\beta_{21}$ & 1.00 & 1.20 & 0.14 & 0.92 & 1.47 & 9.00 & 0.00 \\ 
			$\beta_{21}$ & 1.00 & 1.21 & 0.13 & 0.96 & 1.48 & 0.00 & 0.00 \\ 
			$\beta_{22}$ & 1.00 & 1.02 & 0.18 & 0.66 & 1.34 & 58.00 & 0.00 \\ 
			$\beta_{22}$ & 1.00 & 1.01 & 0.18 & 0.72 & 1.34 & 0.00 & 0.00 \\ 
			\bottomrule
		\end{tabular}
	\end{table}

	\section{Analysis of diabetes data} \label{sec:realdata}
	
	As detailed in Section \ref{sec:data}, we aim to analyse prevalence and mortality risk for MSOAs using prevalence information for GP practices. The dataset comprises 130 GP practices, and the 95 MSOAs residents of which were among the largest contributors to the selected GPs populations.
	
	To analyse the dataset we ran 4 chains each with 2500 iterations of HMC and 50\% warm-up using RStan. The models we considered are the GMCAR-MM and the MCAR-MM both including and excluding covariates.  
	
	\begin{table} 
		\caption{Summary statistics for the estimated parameter for the four models on the real data (`-C' stands for models with covariates). $ \beta_{j1} $: South Asian population coefficient, $ \beta_{j2} $: IMD coefficient for outcome $ j=1,2 $. Results from 4 chains of 2500 iterations and 50\% warm-up with Rstan.}
		\label{tab:est_pars}
		\centering
		\begin{tabular}{rlrrrrrrrr}
			\toprule
			&  & mean & se\_mean & sd & 2.5\% & 5\% & 95\% & 97.5\% & Rhat \\ 
			\midrule
			\multirow{2}{*}{$\alpha_1$} & GMCAR & 0.43 & 0.01 & 0.27 & 0.02 & 0.04 & 0.90 & 0.94 & 1.00 \\ 
			& GMCAR-C & 0.43 & 0.01 & 0.27 & 0.02 & 0.03 & 0.90 & 0.95 & 1.00 \\ 
			\hline
			\multirow{2}{*}{$\alpha_2$} & GMCAR & 0.97 & 0.00 & 0.03 & 0.88 & 0.91 & 1.00 & 1.00 & 1.01 \\ 
			& GMCAR-C & 0.36 & 0.00 & 0.24 & 0.01 & 0.03 & 0.81 & 0.88 & 1.00 \\
			\hline 
			\multirow{2}{*}{$\alpha$} & MCAR & 0.96 & 0.00 & 0.04 & 0.86 & 0.89 & 1.00 & 1.00 & 1.01 \\ 
			& MCAR-C & 0.33 & 0.01 & 0.23 & 0.01 & 0.03 & 0.76 & 0.83 & 1.00 \\ 
			\hline
			\multirow{4}{*}{$\tau_1$} & GMCAR & 5.85 & 0.05 & 2.83 & 1.85 & 2.19 & 11.19 & 12.41 & 1.00 \\ 
			& GMCAR-C & 6.07 & 0.05 & 2.80 & 1.91 & 2.31 & 11.36 & 12.58 & 1.00 \\ 
			& MCAR & 6.94 & 0.06 & 2.91 & 2.44 & 2.94 & 12.24 & 13.73 & 1.00 \\ 
			& MCAR-C & 5.85 & 0.04 & 2.82 & 1.90 & 2.20 & 11.13 & 12.54 & 1.00 \\ 
			\hline
			\multirow{4}{*}{$\tau_2$} & GMCAR & 3.56 & 0.04 & 1.24 & 1.58 & 1.81 & 5.79 & 6.42 & 1.01 \\ 
			& GMCAR-C & 9.75 & 0.04 & 3.04 & 4.66 & 5.26 & 15.18 & 16.39 & 1.00 \\ 
			& MCAR & 3.35 & 0.05 & 1.22 & 1.51 & 1.69 & 5.59 & 6.19 & 1.00 \\ 
			& MCAR-C & 9.19 & 0.04 & 3.06 & 4.02 & 4.74 & 14.66 & 16.13 & 1.00 \\
			\hline 
			\multirow{4}{*}{$\eta_0$} & GMCAR & -0.14 & 0.02 & 0.48 & -1.07 & -0.92 & 0.68 & 0.82 & 1.01 \\ 
			& GMCAR-C & 0.24 & 0.04 & 1.09 & -1.88 & -1.56 & 1.99 & 2.31 & 1.00 \\ 
			& MCAR & 0.54 & 0.01 & 0.26 & 0.04 & 0.13 & 0.96 & 1.05 & 1.01 \\ 
			& MCAR-C & 0.31 & 0.02 & 0.90 & -1.51 & -1.22 & 1.71 & 2.00 & 1.00 \\
			\hline 
			\multirow{2}{*}{$\eta_1$} & GMCAR & 0.18 & 0.00 & 0.11 & -0.04 & -0.01 & 0.35 & 0.39 & 1.01 \\ 
			& GMCAR-C & 0.04 & 0.02 & 0.44 & -0.82 & -0.69 & 0.76 & 0.90 & 1.00 \\
			\hline 
			\multirow{4}{*}{$\psi_1$} & GMCAR & 20.43 & 0.16 & 11.30 & 6.57 & 7.57 & 42.21 & 49.32 & 1.00 \\ 
			& GMCAR-C & 20.60 & 0.14 & 11.22 & 6.71 & 7.82 & 42.51 & 49.92 & 1.00 \\ 
			& MCAR & 18.33 & 0.18 & 10.65 & 5.85 & 6.82 & 39.41 & 46.29 & 1.00 \\ 
			& MCAR-C & 19.87 & 0.16 & 11.17 & 6.37 & 7.33 & 41.04 & 48.01 & 1.00 \\ 
			\hline
			\multirow{4}{*}{$\psi_2$} & GMCAR & 16.56 & 0.04 & 2.70 & 11.98 & 12.57 & 21.35 & 22.48 & 1.00 \\ 
			& GMCAR-C & 17.80 & 0.03 & 2.66 & 12.99 & 13.80 & 22.33 & 23.77 & 1.00 \\ 
			& MCAR & 16.69 & 0.05 & 2.65 & 12.03 & 12.68 & 21.29 & 22.40 & 1.00 \\ 
			& MCAR-C & 17.84 & 0.03 & 2.73 & 13.03 & 13.71 & 22.54 & 23.66 & 1.00 \\ 
			\hline
			\multirow{4}{*}{$\gamma_1$} & GMCAR & 0.19 & 0.01 & 0.16 & -0.09 & -0.04 & 0.49 & 0.57 & 1.02 \\ 
			& GMCAR-C & -0.35 & 0.00 & 0.17 & -0.70 & -0.63 & -0.07 & -0.01 & 1.00 \\ 
			& MCAR & 0.12 & 0.01 & 0.20 & -0.27 & -0.19 & 0.46 & 0.57 & 1.01 \\ 
			& MCAR-C & -0.36 & 0.00 & 0.15 & -0.67 & -0.62 & -0.12 & -0.07 & 1.00 \\ 
			\hline
			\multirow{4}{*}{$\gamma_2$} & GMCAR & 0.12 & 0.01 & 0.18 & -0.23 & -0.14 & 0.43 & 0.51 & 1.04 \\ 
			& GMCAR-C & -0.63 & 0.00 & 0.08 & -0.78 & -0.76 & -0.50 & -0.48 & 1.00 \\ 
			& MCAR & 0.00 & 0.02 & 0.19 & -0.41 & -0.32 & 0.31 & 0.41 & 1.02 \\ 
			& MCAR-C & -0.63 & 0.00 & 0.08 & -0.79 & -0.76 & -0.50 & -0.48 & 1.00 \\ 
			\hline
			\multirow{2}{*}{$\beta_{11}$} & GMCAR-C & 0.51 & 0.00 & 0.23 & 0.04 & 0.12 & 0.90 & 0.96 & 1.00 \\ 
			& MCAR-C & 0.52 & 0.00 & 0.21 & 0.11 & 0.17 & 0.87 & 0.93 & 1.00 \\ 
			\hline
			\multirow{2}{*}{$\beta_{12}$} & GMCAR-C & 0.65 & 0.00 & 0.29 & 0.09 & 0.18 & 1.12 & 1.20 & 1.00 \\ 
			& MCAR-C & 0.67 & 0.00 & 0.27 & 0.15 & 0.24 & 1.10 & 1.19 & 1.00 \\ 
			\hline
			\multirow{2}{*}{$\beta_{21}$} & GMCAR-C & 0.86 & 0.00 & 0.10 & 0.67 & 0.70 & 1.02 & 1.05 & 1.00 \\ 
			& MCAR-C & 0.86 & 0.00 & 0.10 & 0.66 & 0.70 & 1.02 & 1.06 & 1.00 \\ 
			\hline
			\multirow{2}{*}{$\beta_{22}$} & GMCAR-C & 0.83 & 0.00 & 0.13 & 0.57 & 0.61 & 1.04 & 1.07 & 1.00 \\ 
			& MCAR-C & 0.82 & 0.00 & 0.13 & 0.57 & 0.61 & 1.04 & 1.08 & 1.00 \\
			\bottomrule
		\end{tabular}
	\end{table}
	
	Table \ref{tab:est_pars} reports the posterior parameter estimates.
	Consider the \textit{bridging} parameters $\eta_0$ and $\eta_1$: former accounts for \textit{within area} correlation between spatial effects for different outcomes, and the latter represents \textit{between area} association (i.e a form of spatial lag effect). Only $\eta_0$ for the MCAR without covariates shows an entirely positive 95\%  credible interval (0.04, 1.05). For the GMCAR models, $\eta_1$ for the GMCAR without covariates is close to significance with a 90\% credible interval spanning between -0.01 and 0.35. So the association between outcomes remains positive but shows up as via a neighbourhood feedback effect with regards to surrounding areas.
	Posterior samples for the coefficients $\bm{\beta}$ on the covariates, show a strong positive impact, as expected, for both the proportion of South Asian population and the deprivation index IMD, in accounting for both prevalence and mortality variations.
	For mortality, impacts of covariates are significant, albeit smaller than those for prevalence.

	\begin{table}
		\caption{Fit measures summary for all four models for Mortality (95 areas)}
		\label{tab:dev1}
		\centering
		\begin{tabular}{lcccccc}
			\toprule
			& $ D(\bar{\theta}) $ & $ \overline{D(\theta)} $ & $ p_D $ & \textit{DIC} & $ looic $  & \textit{Mean(TAP)} \\
			\midrule
			GMCAR & 78.12 &94.64 &16.52 &111.16 & 412.9 &  0.63\\
			GMCAR (covariates) &74.68 & 93.78 &19.09 &112.87 & 413.7 &0.63\\
			MCAR & 83.40 &98.15 &14.75 & 112.9 &418.0 & 0.56\\
			MCAR (covariates) &78.91 &95.17 &16.26 & 111.43&412.6&0.62  \\
			\bottomrule
		\end{tabular}
	\end{table}
	
	\begin{table}
		\caption{Fit measures summary for all four models for Prevalence (130 GP practices)}
		\label{tab:dev2}
		\centering
		\begin{tabular}{lcccccc}
			\toprule
			& $ D(\bar{\theta}) $ & $ \overline{D(\theta)} $ & $ p_D $ & \textit{DIC} & $ looic $  & \textit{Mean(TAP)}\\
			\midrule
			GMCAR & 121.30 &141.14 &19.84 &160.98 &  1524.2 &0.33 \\
			GMCAR (covariates) &131.62 &142.45 & 10.83&153.28& 1513.0& 0.32\\
			MCAR & 121.23 & 141.46&20.23 &161.69 & 1523.6 & 0.33\\
			MCAR (covariates) &131.12 & 142.34&11.22 &153.56 &1509.9& 0.31\\
			\bottomrule
		\end{tabular}
	\end{table}
	
	\begin{table}
		\caption{Comparison of the mean and standard error of the \textit{elpd} differences: $ \widehat{elpd}^A_{loo} - \widehat{elpd}^B_{loo} $. Positive value indicates better predictive performance for model A, negative values indicate better performance for model B (`-C' stands for models with covariates).}
		\label{tab:loo_comp_nocovVScov}
		\centering
		\begin{tabular}{lr}
			\centering
			\begin{tabular}{llcc}
				\toprule
				\multicolumn{4}{c}{Mortality} \\
				\hline 
				\rule{0pt}{3ex}	Model A & Model B	& Mean  & \textit{se} \\
				\midrule
				GMCAR & MCAR & 2.5  & 1.2 \\  
				GMCAR-C & MCAR-C &-0.6     &  0.4   \\
				GMCAR & GMCAR-C & 0.4   &    1.5 \\  
				MCAR & MCAR-C & -2.3    &   1.3  \\
				\bottomrule
			\end{tabular}
			&
			\begin{tabular}{llcc}
				\toprule
				\multicolumn{4}{c}{Prevalence} \\
				\hline
				\rule{0pt}{3ex} 	Model A & Model B	& Mean  & \textit{se} \\
				\midrule
				GMCAR & MCAR & -0.3    &   0.6 \\
				GMCAR-C & MCAR-C & -2.5    &     1.7 \\
				GMCAR & GMCAR-C & -5.6  &     4.3 \\  
				MCAR & MCAR-C & -8.4   &   2.5   \\
				\bottomrule
			\end{tabular}
		\end{tabular}
	\end{table}
	
	Tables \ref{tab:dev1} and \ref{tab:dev2} compare model fit using the DIC and LOO criteria. For the latter we report $looic = -2 \widehat{elpd}_{loo}$ so that outputs can be compared on the deviance or AIC scale \citep{Vehtari2017}. All models fit the data adequately, but there are important differences in the results when covariates are included in the models. For prevalence, inclusion of covariates leads to reduced DIC and LOO-IC. It can be seen that there are relatively small differences on these criteria between comparable models (i.e. models with covariates and models without covariates). The average posterior deviances are approximately close to the number of observations, suggesting a good fit of the model to the data. We compared LOO for GMCAR-MM and MCAR-MM models using differences in \textit{elpd} as described in Section \ref{sec:mod_comp}. This shows that the GMCAR-MM mortality model is preferable to the MCAR-MM one, showing a difference of 2.5 (on the \textit{elpd} scale) and a standard error of 1.2. Also, the addition of covariates shows a significant improvement for the MCAR-MM prevalence model with an average \textit{elpd} difference of 8.4 (standard error 2.5). Other model comparisons in Table \ref{tab:loo_comp_nocovVScov} do not show significant differences in leave-one-out cross validation prediction.
	
	\begin{table}
		\caption{Proportion of tail area probabilities in different intervals for the four models and both outcomes (`-C' stands for models with covariates).}
		\label{tab:tail_p}
		\centering
		\begin{tabular}{lcccc}
			\toprule
			& \multicolumn{2}{c}{Mortality} & \multicolumn{2}{c}{Prevalence}  \\
			& $ \left[0, 0.05\right] \cup \left[0.95, 1\right]$ & $ \left[0, 0.1\right] \cup \left[0.90, 1\right]$ & $ \left[0, 0.05\right] \cup \left[0.95, 1\right]$ & $ \left[0, 0.1\right] \cup \left[0.90, 1\right]$ \\
			\midrule
			GMCAR & 0.042 & 0.08 & 0.015 &  0.023 \\
			GMCAR-C & 0.032 & 0.074& 0.015 &0.023 \\
			MCAR &0.042 &0.084 &0.015 & 0.015 \\
			MCAR-C &0.042 &0.074 &0.015 & 0.023\\
			\bottomrule
		\end{tabular}
	\end{table}
	
\begin{figure}
		\centering
		\caption{Densities for mortality $ y $ and replicated data at each replication $ y_{rep} $}
		\label{fig:dens_rep1}
		\subfloat[]{\includegraphics[width=7.5cm]{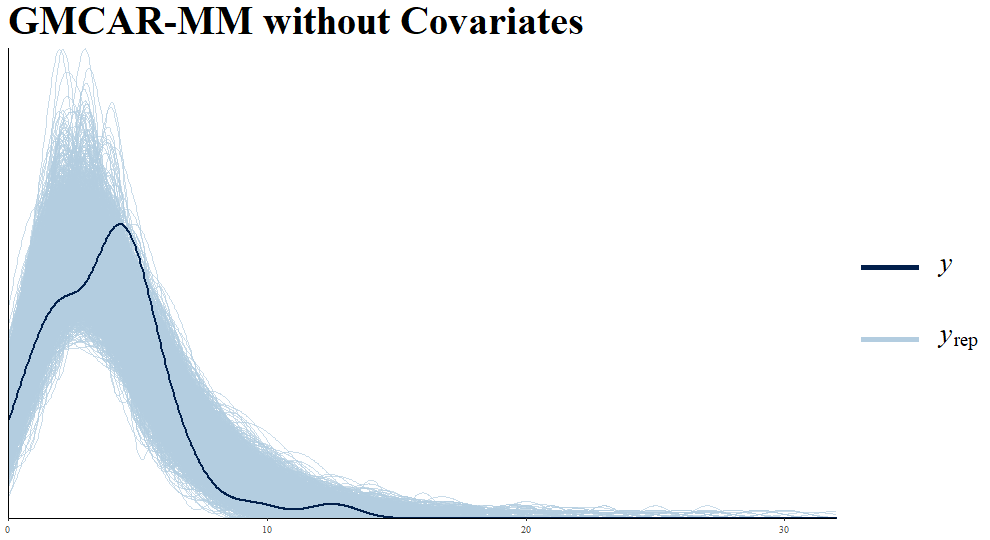}}
		\subfloat[]{\includegraphics[width=7.5cm]{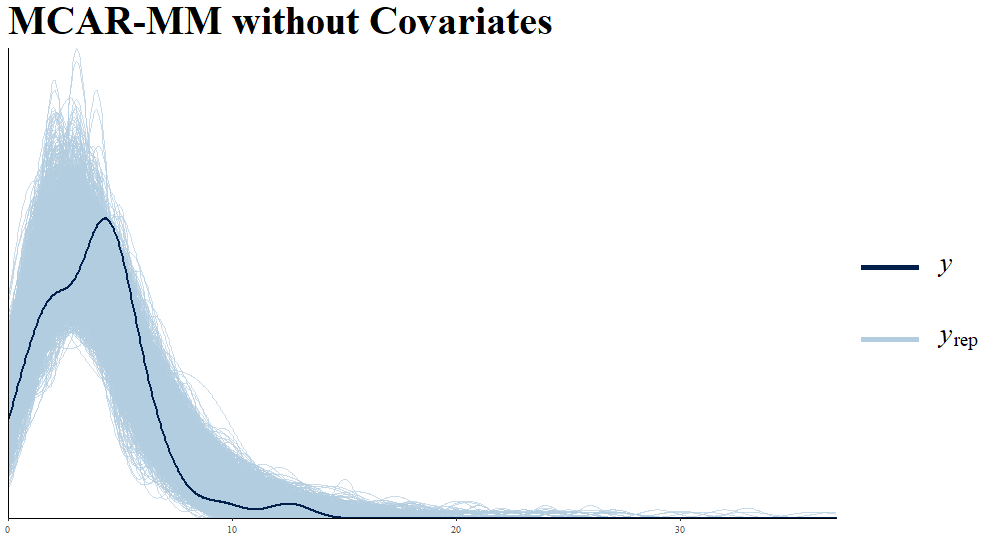}}
		\\
		\subfloat[]{\includegraphics[width=7.5cm]{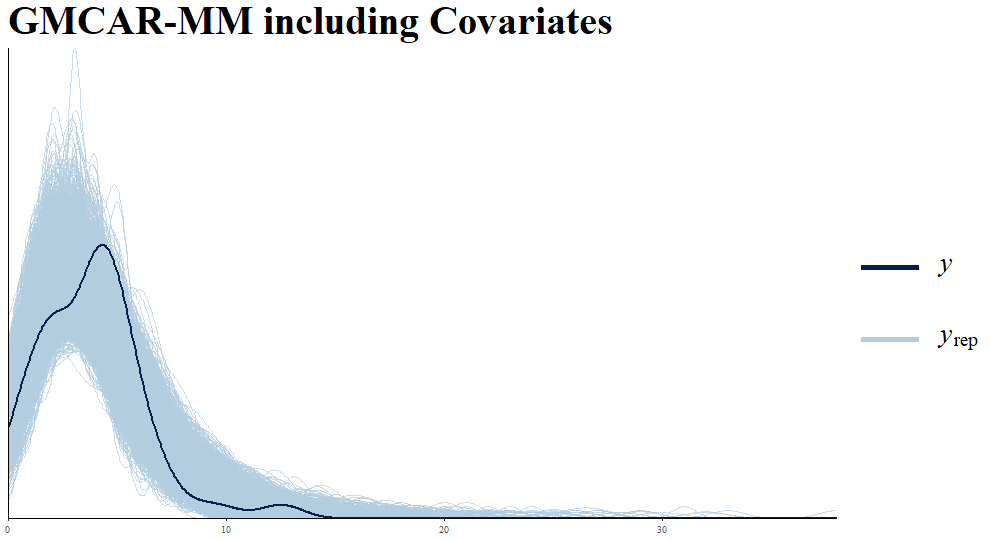}}
		\subfloat[]{\includegraphics[width=7.5cm]{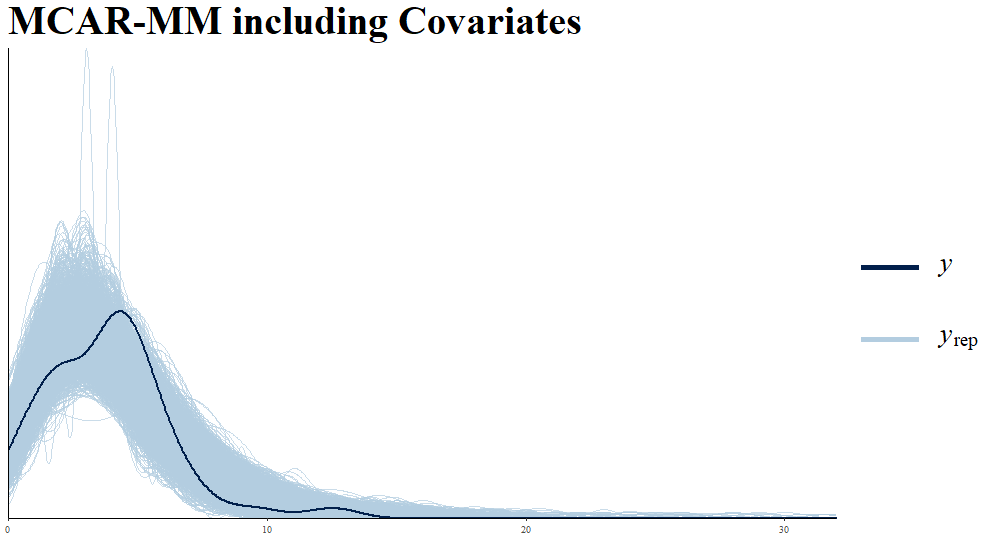}}
	\end{figure}

\begin{figure}
		\centering
		\caption{Densities for prevalence $ y $ and replicated data at each replication $ y_{rep} $}
		\label{fig:dens_rep2}
		\subfloat[]{\includegraphics[width=7.5cm]{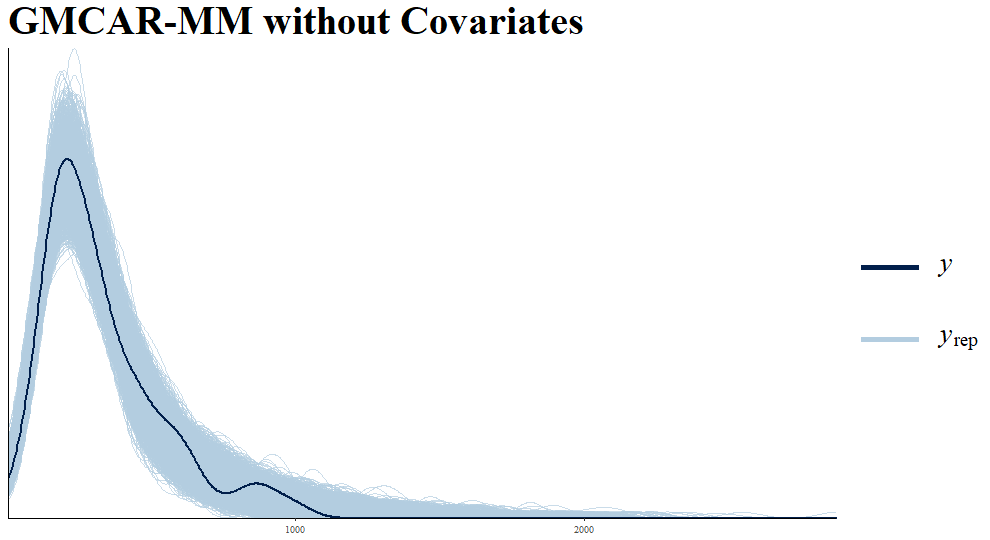}}
		\subfloat[]{\includegraphics[width=7.5cm]{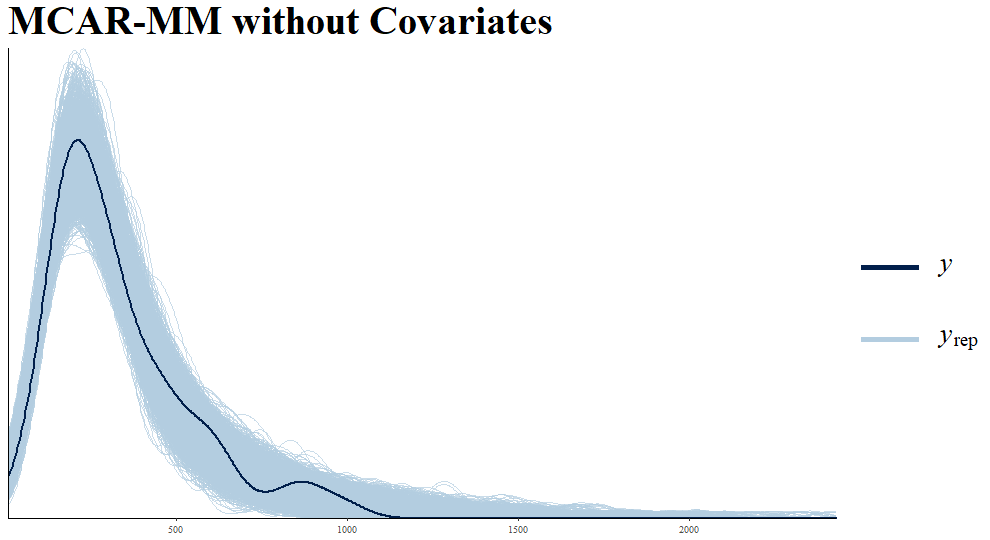}}
		\\
		\subfloat[]{\includegraphics[width=7.5cm]{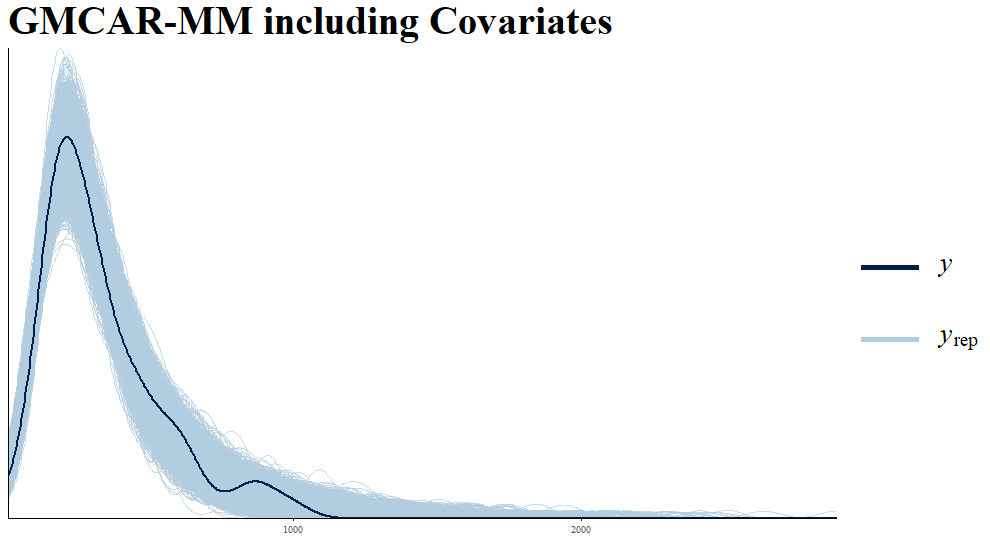}}
		\subfloat[]{\includegraphics[width=7.5cm]{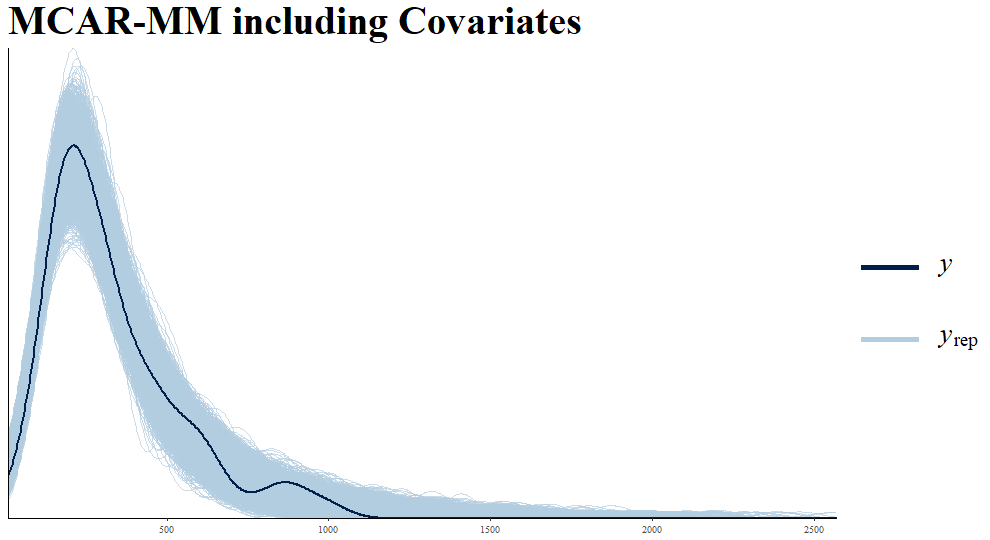}}
	\end{figure}

	Again, Tables \ref{tab:dev1} and \ref{tab:dev2} show that the estimated posterior mean values of TAP never exceed 0.63 and are never below 0.31, as expected with adequately fitting models. Additionally, in Table \ref{tab:tail_p} we can clearly see that it is very unlikely for the replicated values across all models to be sampled systematically over or under the dataset values. Figures \ref{fig:dens_rep1} and \ref{fig:dens_rep2} display densities for the replicated data at each iteration with the actual data.
	
	\begin{figure}
		\centering
		\caption{Estimated mortality relative risks (posterior means of $ \bm{\rho}_1 $) for the area of interest according to the four models.}
		\label{fig:map_1}
		\subfloat[]{\includegraphics[width=15cm]{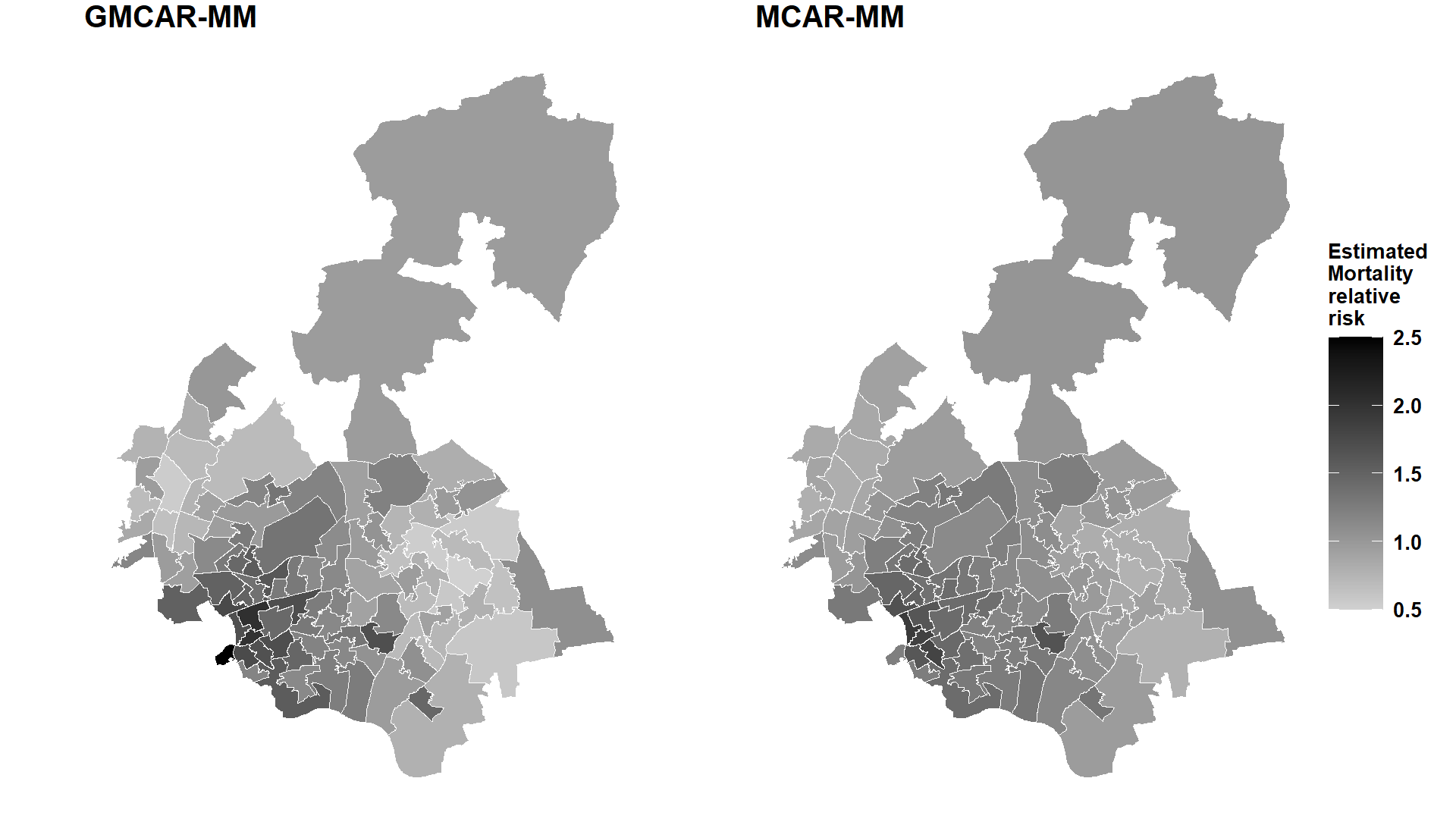}} \\
		\subfloat[]{\includegraphics[width=15cm]{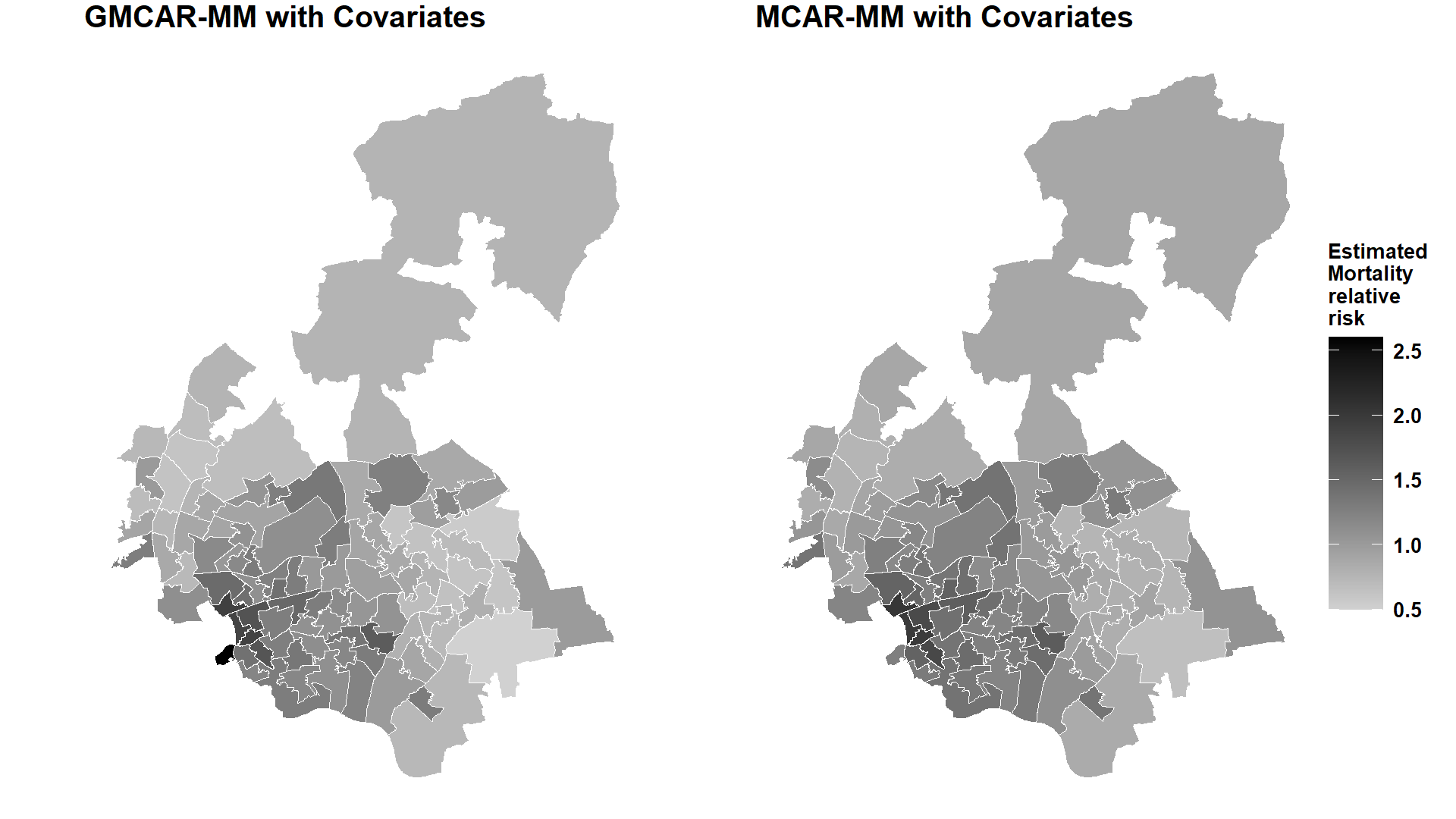}}
	\end{figure}
	
	\begin{figure}
		\centering
		\caption{Estimated prevalence relative risks for the area of interest (posterior means of $ \bm{\zeta}_2 $) according to the four models.}
		\label{fig:map_2}
		\subfloat[]{\includegraphics[width=15cm]{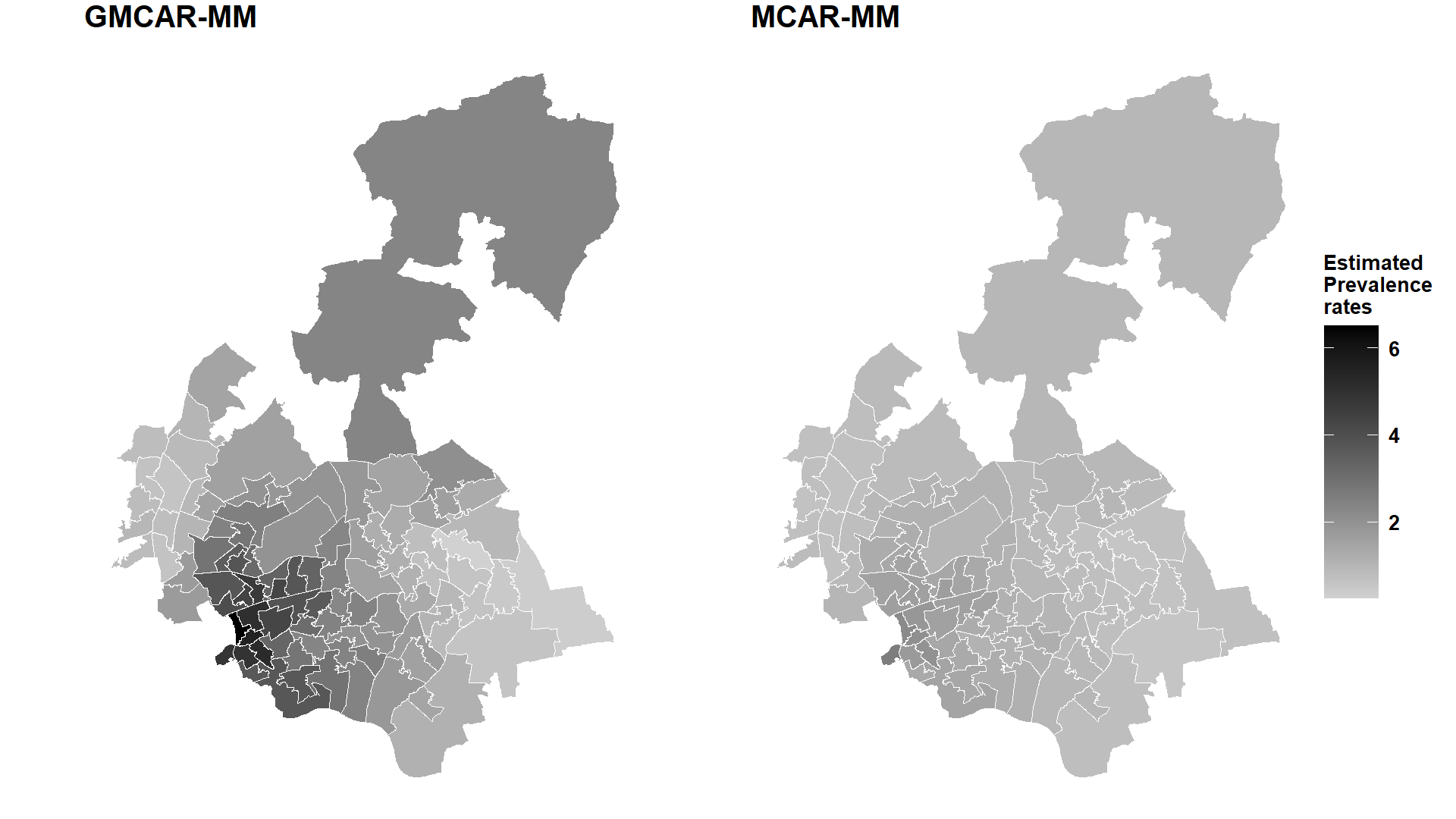}} \\
		\subfloat[]{\includegraphics[width=15cm]{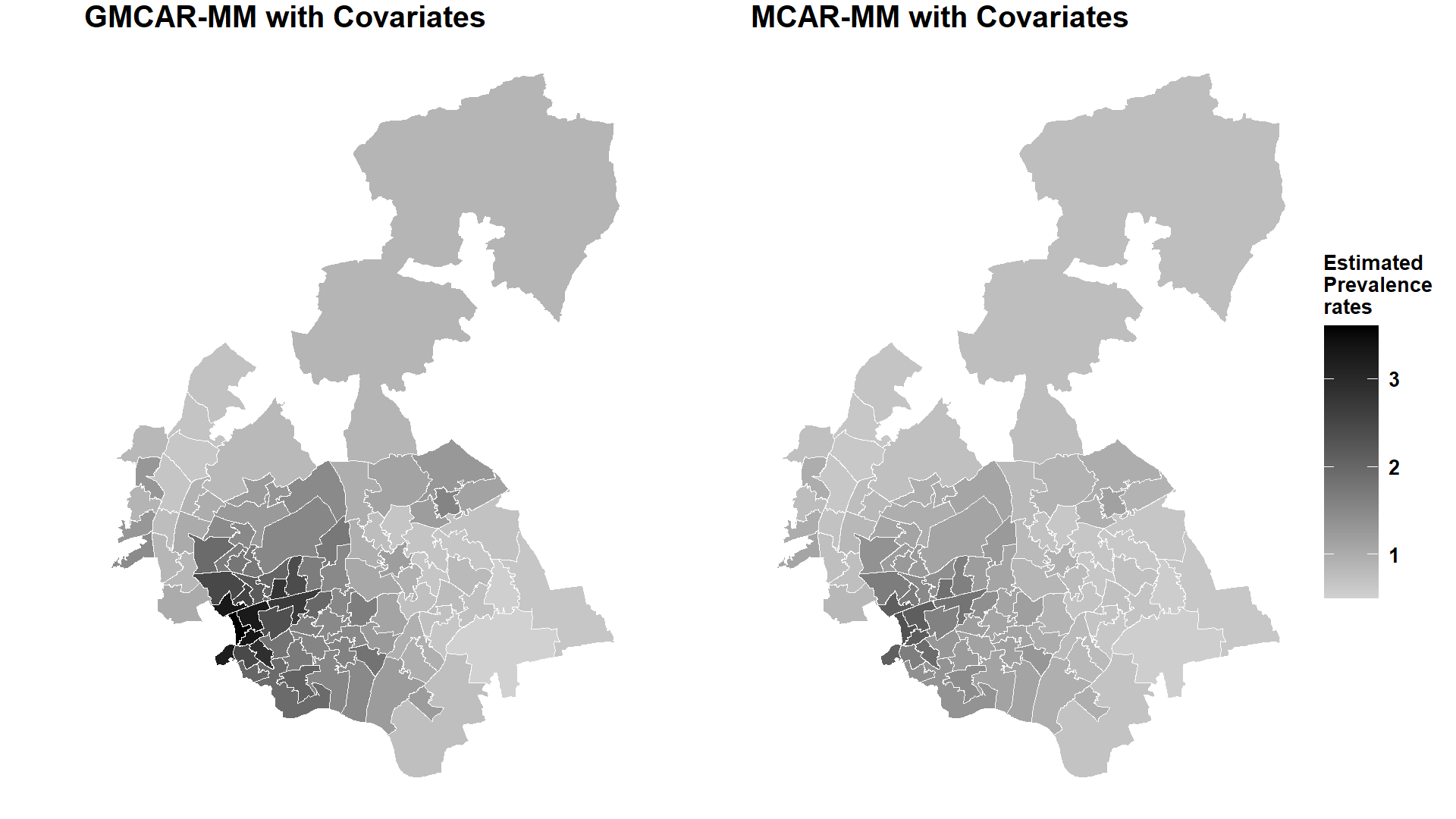}}
	\end{figure}
	
	
	\begin{table}
		\caption{Pearson correlation between areal relative risks ($\bm{\rho}$ and $\bm{\zeta}$) at each iteration of the HMC on the ONEL data (4 chains of 2500 iterations and 50\% warm-up with Rstan).}
		\label{tab:corr_risk}
		\centering
		\begin{tabular}{lcccccccc}
			\toprule
			& mean & se\_mean & sd & 2.5\% & 5\% & 95\% & 97.5\% & Rhat\\
			\midrule
			GMCAR & 0.43 & 0.01 & 0.25 & -0.07 & -0.01 & 0.81 & 0.85 & 1 \\ 
			GMCAR-C & 0.61 & 0.00 & 0.18 & 0.19 & 0.26 & 0.85 & 0.87 & 1 \\ 
			MCAR & 0.70 & 0.00 & 0.16 & 0.30 & 0.39 & 0.90 & 0.92 & 1 \\ 
			MCAR-C & 0.65 & 0.00 & 0.17 & 0.25 & 0.33 & 0.87 & 0.89 & 1 \\ 
			\bottomrule
		\end{tabular}
	\end{table}
	
	Posterior Pearson's correlation between relative risks is shown in Table \ref{tab:corr_risk}. As expected the two outcomes display a positive correlation that is increased by the addition of covariates to the model in the case of the GMCAR, and slightly decreased, though remaining rather high, for the MCAR.

\begin{figure}[ht]
		\caption{Mapping of areal risk categories (Section \ref{sec:risk_clustering}) for \textit{mortality} and the four models ($ T_R = 1 $ see Eq. \ref{eq:rr_prob} and $ T_P = 0.9 $)}
		\label{fig:clust_adj_mort}
		\subfloat[]{\includegraphics[width=15cm]{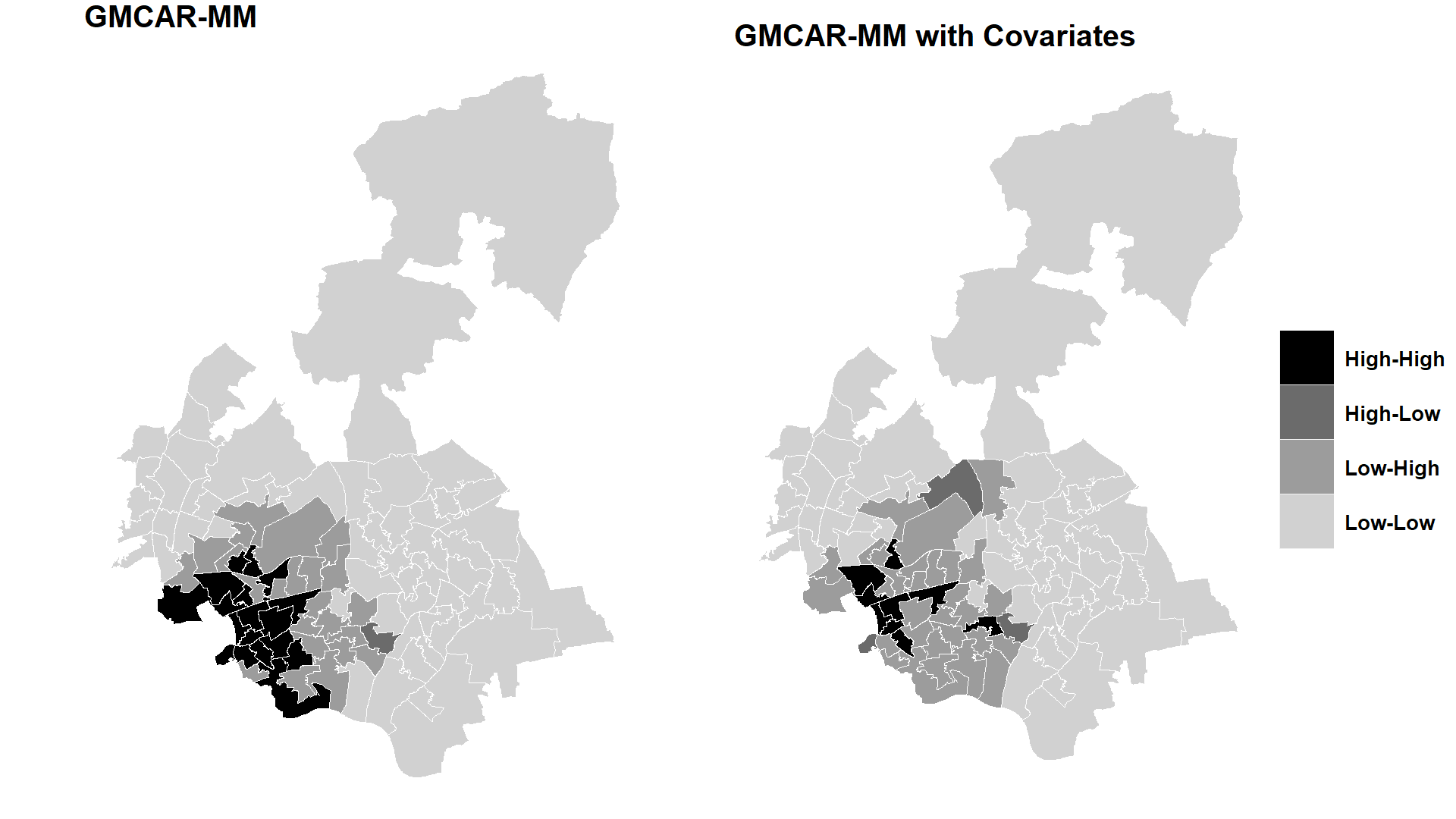}} \\
		\subfloat[]{\includegraphics[width=15cm]{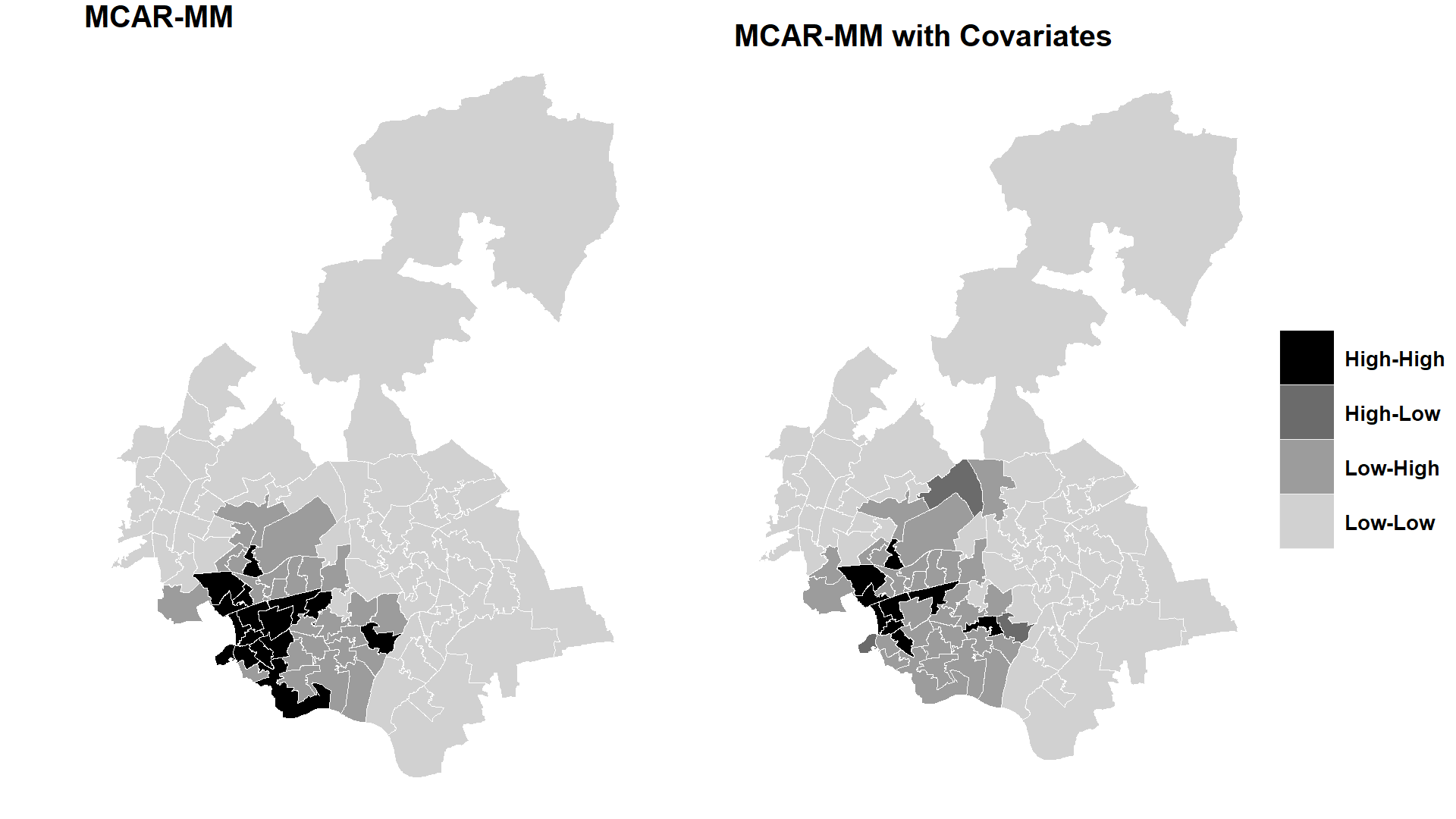}}
	\end{figure}

\begin{figure}[ht]
		\caption{Mapping of areal risk categories (Section \ref{sec:risk_clustering}) for \textit{prevalence} and the four models ($ T_R = 1 $ see Eq. \ref{eq:rr_prob} and $ T_P = 0.9 $)}
		\label{fig:clust_adj_prev}
		\subfloat[]{\includegraphics[width=15cm]{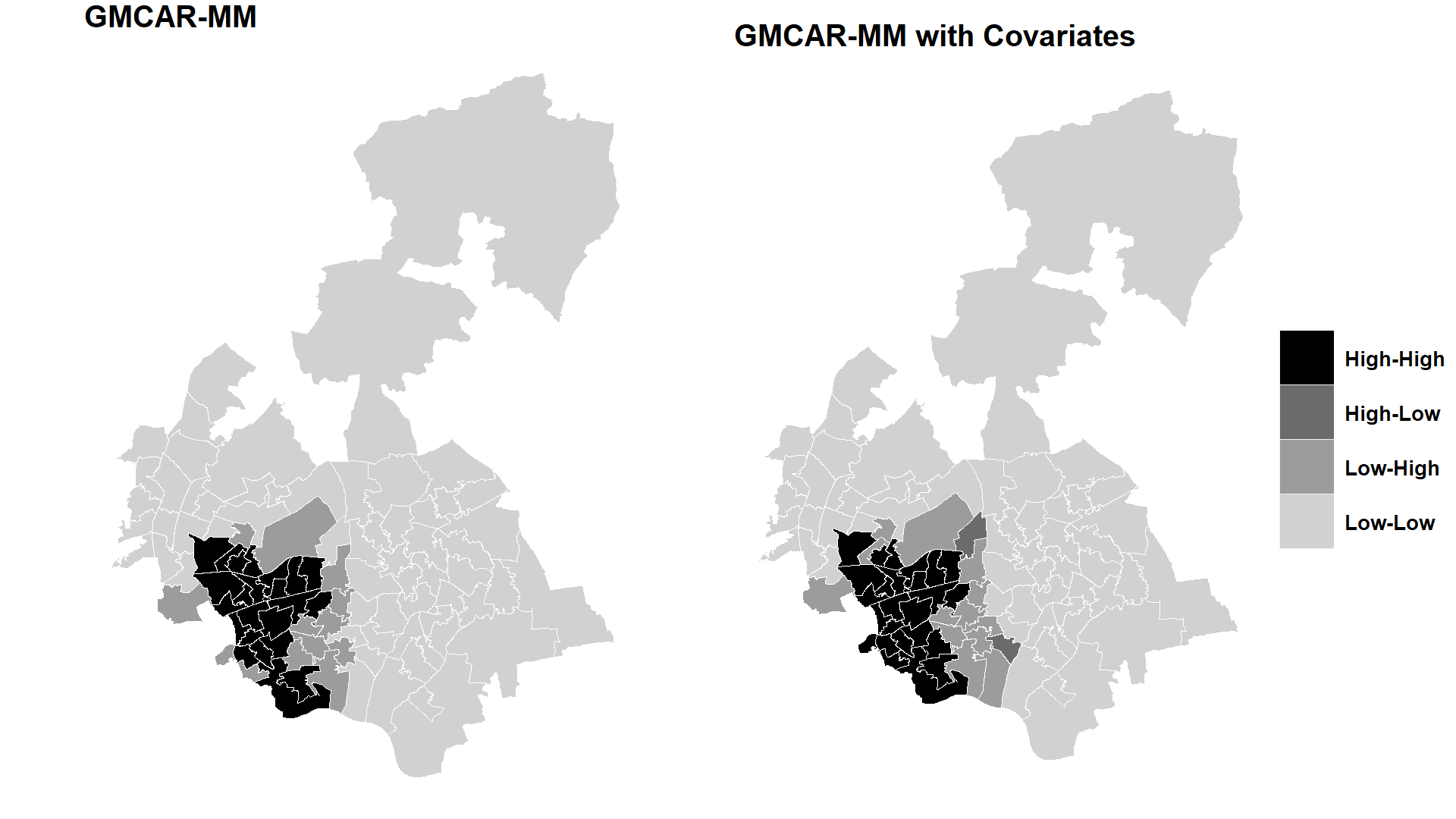}} \\
		\subfloat[]{\includegraphics[width=15cm]{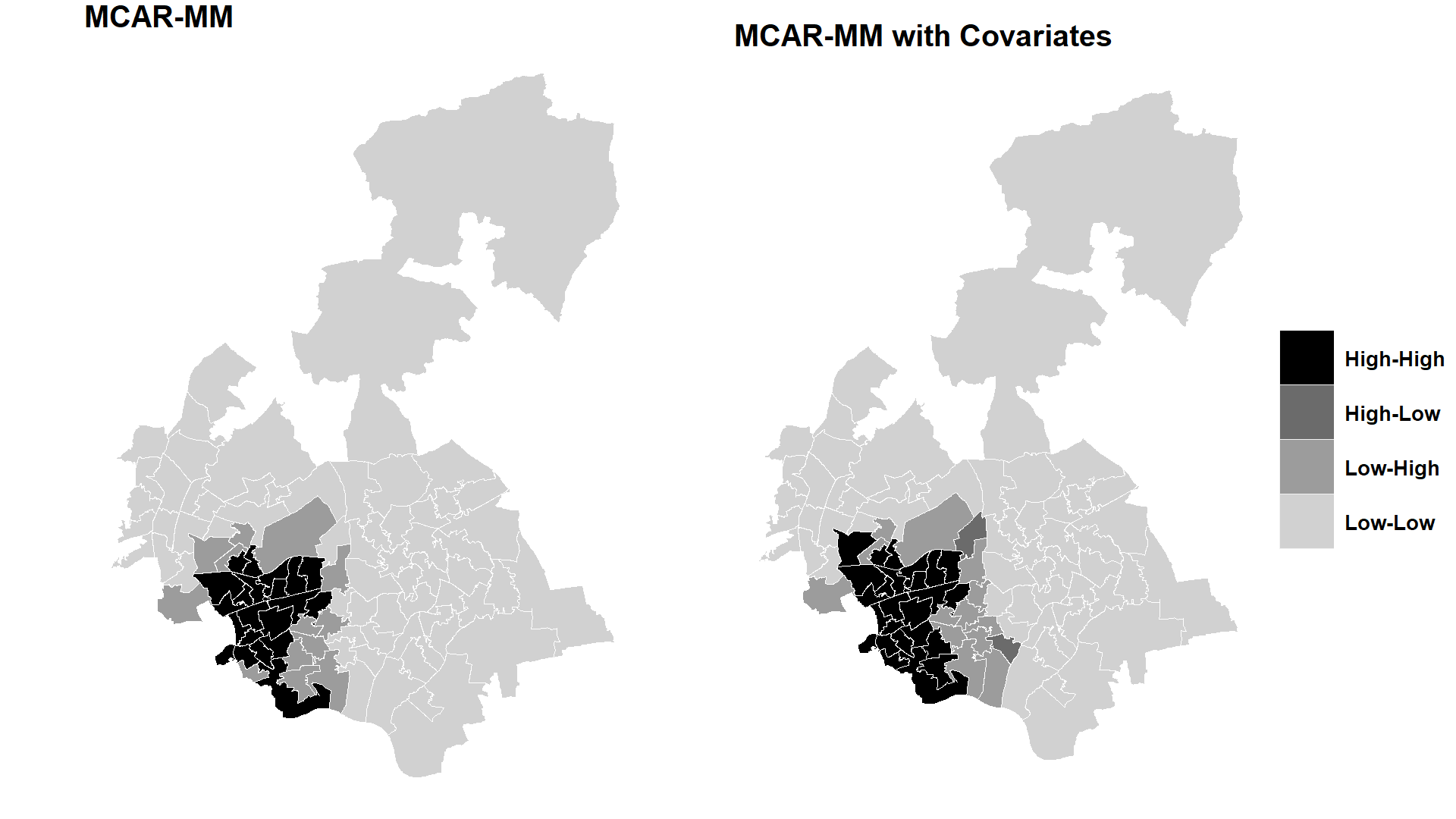}}
	\end{figure}

	Figure \ref{fig:map_1} shows maps of posterior means for the estimated mortality relative rates in MSOAs, i.e. $ \bm{\rho_1} $ of equation (\ref{eq:full}). Figure \ref{fig:map_2} plots the posterior means of the estimated relative risks for prevalence in areas (as against the GP practices), i.e. $ \bm{\zeta}_2 $ as in equation (\ref{eq:rr_membToArea}). We analyse these maps further with clustering analysis: Figures \ref{fig:clust_adj_mort} and \ref{fig:clust_adj_prev} show the areas belonging to the four risk cluster categories. When an area (MSOA) is drawn in black, it means that it belongs to the HH category and therefore it is surrounded by other areas with a global average of high relative risk, while being itself of high relative risk, therefore lying at the centre of a high risk cluster. Most MSOAs tend to lie in the LL category (see also Tables \ref{tab:clust_mort} and \ref{tab:clust_prev}), with the higher risk ones (HH and HL) concentrated in the south west part of the map, namely areas adjacent to inner London. 
	
	For mortality, as noted above the addition of covariates tends to decrease relative risk overall. However, Figure \ref{fig:clust_adj_mort} highlights two MSOAs of particular interest. Redbridge 002 (RB, E02000752) and Barking and Dagenham 009 (BD, E02000010) move from low relative risk mortality in the models without covariates to high relative risk for that same outcome, while prevalence maintains the low risk profile across all models. Both areas have an high mortality count (RB: 7, BD: 6) but comparatively low expected counts (RB: 3.02, BD: 2.63) and both areas are bordering mostly areas with few deaths. Consequently, in the intercept-only models, relative risks for those areas were likely shrunk toward the lower neighbourhood values, while the addition of covariates allows the model to discriminate these areas, and highlight their higher risk.
	
	Additional covariates would help shedding more light on the reason why areas in Redbridge and Barking and Dagenham CCGs report such a high number of deaths by diabetes.	
	
	Similarly, in the case of prevalence (Figure \ref{fig:clust_adj_prev}) Barking and Dagenham 001 and 014 (E02000002 and E02000015) are classified as high risk only when covariates are included. While both areas have a relatively low share of South Asian residents (8.78 and 3.44 \%, global mean of 15.67 \%), they are respectively the $ 8^{th} $ and $ 1^{st} $ most deprived MSOAs in the ONEL area (IMD of 38.5 and 43.2, global mean of 22.56).
	This explains how the inclusion of covariates augments the relative risk, since the effect of deprivation is estimated globally on the ONEL area.
	
\begin{figure}[ht]
		\caption{Mapping of \textit{bivariate within area} risk categories (Section \ref{sec:risk_clustering}) for the four models ($ T_R = 1 $ see Eq. \ref{eq:rr_prob} and $ T_P = 0.9 $)}
		\label{fig:clust_biv}
		\subfloat[]{\includegraphics[width=15cm]{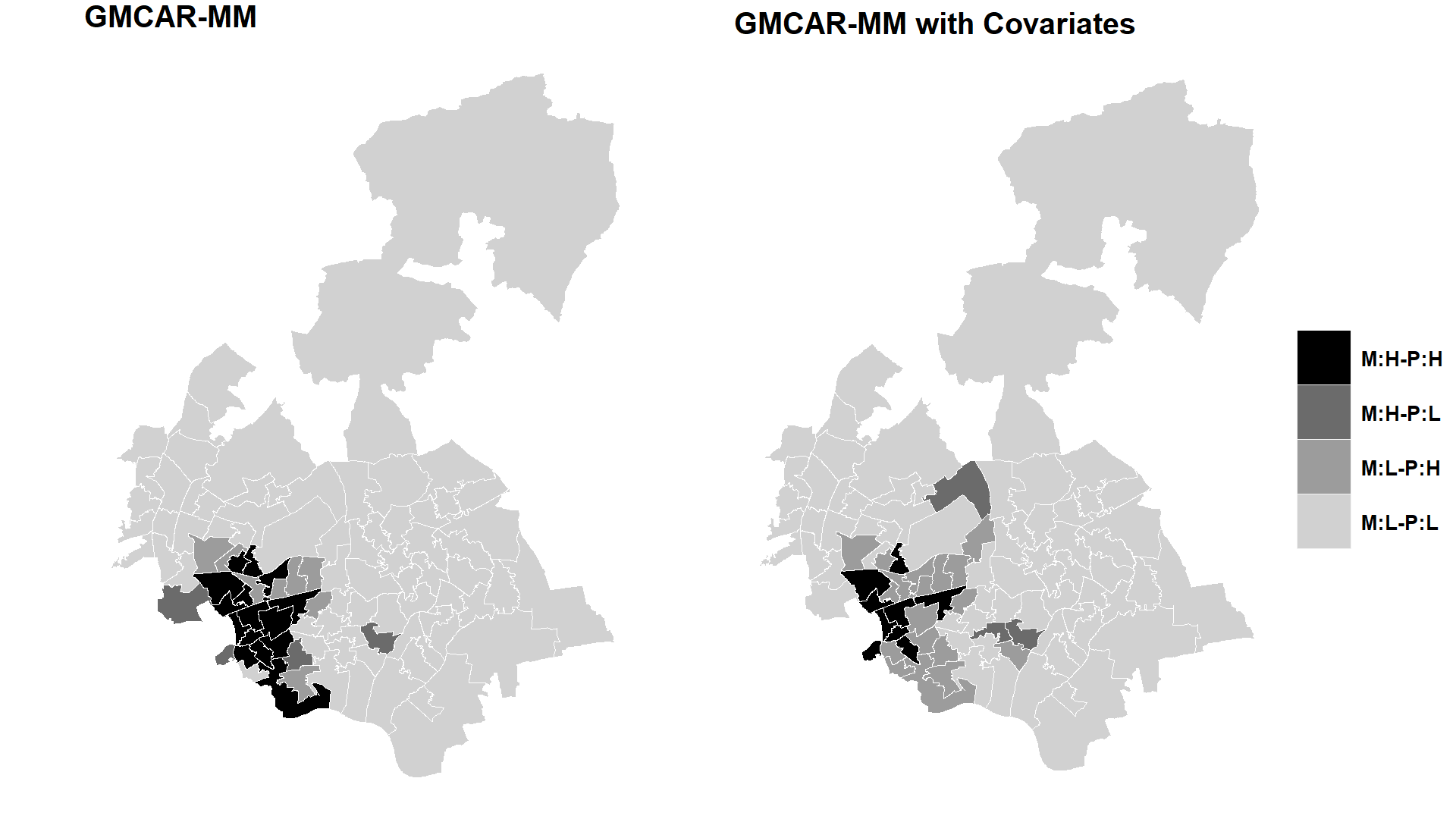}} \\
		\subfloat[]{\includegraphics[width=15cm]{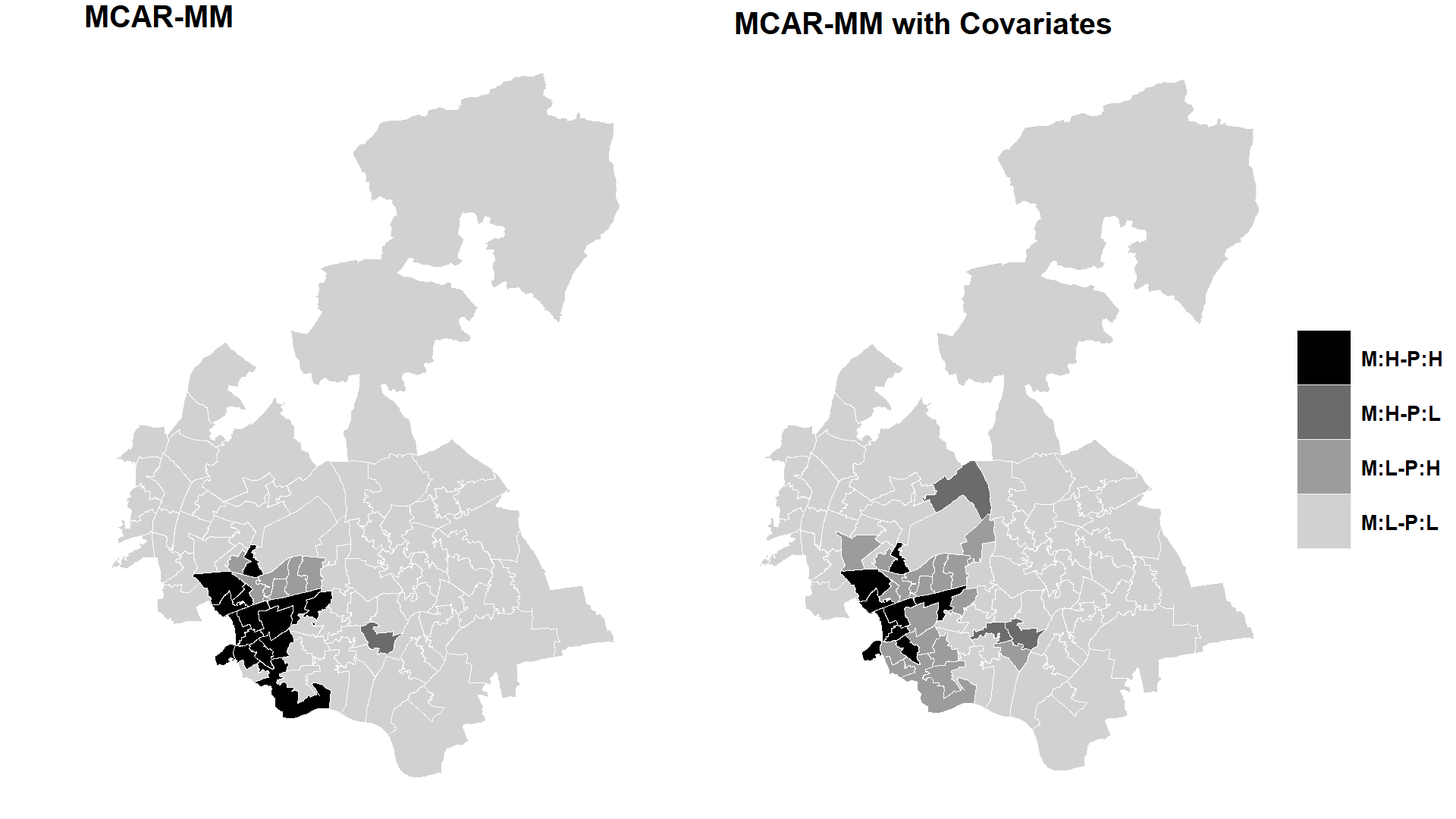}}
	\end{figure}

	Figure \ref{fig:clust_biv} also highlights those same areas just mentioned, by displaying the maps of bivariate risk clustering.
	
	\begin{table}[ht]
		\centering
		\caption{Number of areas belonging to the four clustering categories, arising from comparing an area's probability of high relative risk ($P_H(i, 1)$, Eq. \ref{eq:rr_prob_area}) to its locality probability ($ P_H(\sim i, 1)$, Eq. \ref{eq:rr_prob_neigh}) for \textit{mortality} from the four models}
		\label{tab:clust_mort}
		\begin{tabular}{lr}
			\begin{tabular}{l|c|c} 
				\toprule
				\multicolumn{3}{c}{\textbf{GMCAR-MM}} \\
				\midrule
				& H $P_H(\sim i, 1)$ & L $P_H(\sim i, 1)$ \\
				\hline
				H $P_H(i, 1)$ & 18 & 1 \\
				\hline
				L $P_H(i, 1)$ & 23 & 53 \\
				\bottomrule
			\end{tabular}
			&
			\begin{tabular}{l|c|c}
				\toprule
				\multicolumn{3}{c}{\textbf{GMCAR-MM with Covariates}} \\
				\midrule
				& H $P_H(\sim i, 1)$ & L $P_H(\sim i, 1)$ \\
				\hline
				H $P_H(i, 1)$ & 9 & 3 \\
				\hline
				L $P_H(i, 1)$ & 30 & 53 \\
				\bottomrule
			\end{tabular}
			\\
			\begin{tabular}{l|c|c} 
				\toprule
				\multicolumn{3}{c}{\textbf{MCAR-MM}} \\
				\midrule
				& H $P_H(\sim i, 1)$ & L $P_H(\sim i, 1)$ \\
				\hline
				H $P_H(i, 1)$ & 16 & 0 \\
				\hline
				L $P_H(i, 1)$ & 25 & 54 \\
				\bottomrule
			\end{tabular}
			&
			\begin{tabular}{l|c|c}
				\toprule
				\multicolumn{3}{c}{\textbf{MCAR-MM with Covariates}} \\
				\midrule
				& H $P_H(\sim i, 1)$ & L $P_H(\sim i, 1)$ \\
				\hline
				H $P_H(i, 1)$ & 19 & 1 \\
				\hline
				L $P_H(i, 1)$ & 23 & 52 \\
				\bottomrule
			\end{tabular}
		\end{tabular}
	\end{table}
	
	\begin{table}[ht]
		\centering
		\caption{Number of areas belonging to the four clustering categories, arising from comparing an area's probability of high relative risk ($P_H(i, 1)$, Eq. \ref{eq:rr_prob_area}) to its locality probability ($ P_H(\sim i, 1) $, Eq. \ref{eq:rr_prob_neigh}) for \textit{prevalence} from the four models}
		\label{tab:clust_prev}
		\begin{tabular}{lr}
			\begin{tabular}{l|c|c} 
				\toprule
				\multicolumn{3}{c}{\textbf{GMCAR-MM}} \\
				\midrule
				& H $P_H(\sim i, 1)$ & L $P_H(\sim i, 1)$ \\
				\hline
				H $P_H(i, 1)$ & 23 & 0 \\
				\hline
				L $P_H(i, 1)$ & 13 & 59 \\
				\bottomrule
			\end{tabular}
			&
			\begin{tabular}{l|c|c}
				\toprule
				\multicolumn{3}{c}{\textbf{GMCAR-MM with Covariates}} \\
				\midrule
				& H $P_H(\sim i, 1)$ & L $P_H(\sim i, 1)$ \\
				\hline
				H $P_H(i, 1)$ & 25 & 2 \\
				\hline
				L $P_H(i, 1)$ & 13 & 55 \\
				\bottomrule
			\end{tabular}
			\\
			\begin{tabular}{l|c|c} 
				\toprule
				\multicolumn{3}{c}{\textbf{MCAR-MM}} \\
				\midrule
				& H $P_H(\sim i, 1)$ & L $P_H(\sim i, 1)$ \\
				\hline
				H $P_H(i, 1)$ & 21 & 0 \\
				\hline
				L $P_H(i, 1)$ & 13 & 61 \\
				\bottomrule
			\end{tabular}
			&
			\begin{tabular}{l|c|c}
				\toprule
				\multicolumn{3}{c}{\textbf{MCAR-MM with Covariates}} \\
				\midrule
				& H $P_H(\sim i, 1)$ & L $P_H(\sim i, 1)$ \\
				\hline
				H $P_H(i, 1)$ & 25 & 2 \\
				\hline
				L $P_H(i, 1)$ & 14 & 54 \\
				\bottomrule
			\end{tabular}
		\end{tabular}
	\end{table}
	
	\begin{table}[ht]
		\centering
		\caption{Number of areas classified according to \textit{bivariate within area} risk categories (Section \ref{sec:risk_clustering}) for the four models ($ T_R = 1 $ see Eq. \ref{eq:rr_prob} and $ T_P = 0.9 $)}
		\label{tab:bivar_clust}
		\begin{tabular}{lr}
			\begin{tabular}{l|l|c|c|c|c} 
				\toprule
				&\multicolumn{5}{c}{\textbf{GMCAR-MM}} \\
				\cline{2-6}
				&	& \multicolumn{4}{c}{\textbf{Prevalence}} \\
				\cline{3-6}
				&  & H-H & H-L & L-H & L-L\\
				\hline
				\parbox[t]{2mm}{\multirow{4}{*}{\rotatebox[origin=c]{90}{\textbf{Mortality}}}} & H-H  & 15 & 0 &3 & 0\\
				\cline{2-6}
				& H-L  & 0 & 0 &0 & 1 \\
				\cline{2-6}
				& L-H  & 8 & 0 &9 &6 \\
				\cline{2-6}
				& L-L  & 0 & 0 & 1 & 52\\
				\bottomrule
			\end{tabular}
			&
			\begin{tabular}{l|l|c|c|c|c} 
				\toprule
				&\multicolumn{5}{c}{\textbf{GMCAR-MM with Covariates}} \\
				\cline{2-6}
				&	& \multicolumn{4}{c}{\textbf{Prevalence}} \\
				\cline{3-6}
				&  & H-H & H-L & L-H & L-L\\
				\hline
				\parbox[t]{2mm}{\multirow{4}{*}{\rotatebox[origin=c]{90}{\textbf{Mortality}}}} & H-H  & 8 & 0 & 1 &  0\\
				\cline{2-6}
				& H-L  & 1 & 0 & 0 & 2 \\
				\cline{2-6}
				& L-H  & 15 & 1 & 10 &4 \\
				\cline{2-6}
				& L-L  & 1 & 1 & 2 & 49\\
				\bottomrule
			\end{tabular}
			\\  & \\
			\begin{tabular}{l|l|c|c|c|c} 
				\toprule
				&\multicolumn{5}{c}{\textbf{MCAR-MM}} \\
				\cline{2-6}
				&	& \multicolumn{4}{c}{\textbf{Prevalence}} \\
				\cline{3-6}
				&  & H-H & H-L & L-H & L-L\\
				\hline
				\parbox[t]{2mm}{\multirow{4}{*}{\rotatebox[origin=c]{90}{\textbf{Mortality}}}} & H-H  & 15 & 0 &0 & 1\\
				\cline{2-6}
				& H-L  & 0 & 0 &0 & 0 \\
				\cline{2-6}
				& L-H  & 6 & 0 &12 &7 \\
				\cline{2-6}
				& L-L  & 0 & 0 & 1& 53\\
				\bottomrule
			\end{tabular}
			&
			\begin{tabular}{l|l|c|c|c|c} 
				\toprule
				&\multicolumn{5}{c}{\textbf{MCAR-MM with Covariates}} \\
				\cline{2-6}
				&	& \multicolumn{4}{c}{\textbf{Prevalence}} \\
				\cline{3-6}
				&  & H-H & H-L & L-H & L-L\\
				\hline
				\parbox[t]{2mm}{\multirow{4}{*}{\rotatebox[origin=c]{90}{\textbf{Mortality}}}} & H-H  & 8 & 0 &7 & 4\\
				\cline{2-6}
				& H-L  & 0 & 0 &0 & 1 \\
				\cline{2-6}
				& L-H  & 0 & 0 & 15 & 8 \\
				\cline{2-6}
				& L-L  & 0 & 0 & 1 & 51\\
				\bottomrule
			\end{tabular}
		\end{tabular}
	\end{table}
	
	When instead we move to bivariate clustering, we can see in Table \ref{tab:bivar_clust} that under all four models most of the within area risk categories tend to agree, i.e. they tend to lie in the top left corner cells (between 7\% and 15\% M:H-P:H) and bottom right (between 75\% and 87 \% M:L-P:L). This shows that almost all high (low) risk areas for prevalence are also high (low) risk areas for mortality. We note that this holds true for the MCAR-MM model as well, which unlike the GMCAR-MM model, does not assume causal precedence.

	\section{Conclusions}\label{sec:concl}
	
	No previous studies have investigated the bivariate spatial relationship between diabetes mortality and prevalence for small areas. Evidence on the small area patterning of diabetes has gained increased importance as diabetes is a recognized risk factor for Covid-19 \parencite{Guo2020}. 
	
	Although recording of diabetes deaths data can be problematic, with possible undercounting, the models we propose show strong impacts of acknowledged area risk factors (area deprivation and ethnic population mix) on both outcomes. In models not including covariates, spatial association and dependence parameters are mostly in line with broader epidemiological knowledge: e.g. a positive association between the two outcomes. 
	
	Analysis of posterior relative risks shows clear clustering of high risk in a relatively few areas in the western part of the region studied, while low risk (and low-low clusters) characterise a higher number of areas. Further exploration of the data with different CAR priors such as the DAGAR prior \citep{Datta2019}, whose $ \rho $ parameter better captures areal correlation, might help explain spatial autocorrelation as well as help with clustering analysis.
	
	The analysis is complicated by the fact that prevalence is not observed directly for areas but for a different set of spatial units: general practitioner catchment areas. We have used a form of multiple membership principle to transfer information between these spatial frameworks. Posterior checks confirm that this approach is successful.
	
	To the best of our knowledge, so far RStan has been limited to univariate CAR models. The approach we have used extends use of RStan to a bivariate CAR application, and can be extended to more than two outcomes using the conditionality principle.

	
	\printbibliography
\end{document}